\documentclass[table, hidelinks]{preprint}
\usepackage[utf8]{inputenc}
\usepackage[T1]{fontenc}
\usepackage{graphicx}
\usepackage{longtable}
\usepackage{wrapfig}
\usepackage{rotating}
\usepackage[normalem]{ulem}
\usepackage{amsmath}
\usepackage{amssymb}
\usepackage{capt-of}
\usepackage{hyperref}
\usepackage{xcolor}
\usepackage{booktabs}
\usepackage[spanish, es-tabla]{babel}

\usepackage{ebgaramond}
\newbox{\myorcidaffilbox}
\sbox{\myorcidaffilbox}{\large\includegraphics[height=1.7ex]{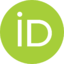}}
\newcommand{\orcidaffil}[1]{%
\href{https://orcid.org/#1}{\usebox{\myorcidaffilbox}}}
\newbox{\myemailbox}
\sbox{\myemailbox}{\large\includegraphics[height=1.7ex]{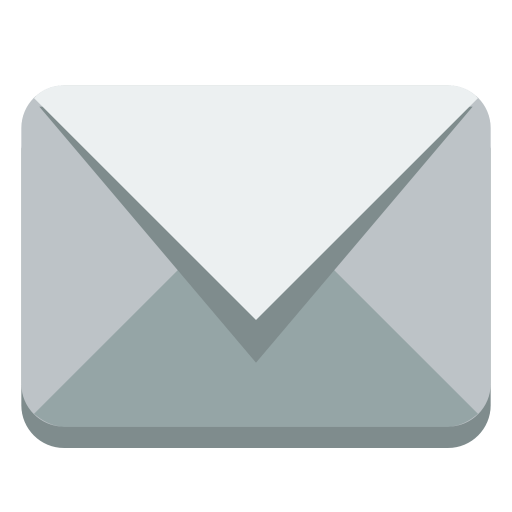}}
\newcommand{\emailicon}[1]{%
\href{mailto:#1?Subject=Re: Sistemas de información de salud en contextos extremos: Uso de teléfonos móviles para combatir el sida en Uganda}{\usebox{\myemailbox}}}
\usepackage{authblk}
\usepackage{orcidlink}
\author[1,2]{Livingstone Njuba, MSc}
\author[3]{Juan E. Gómez-Morantes \orcidaffil{0000-0002-8107-4030}\emailicon{je.gomezm@javeriana.edu.co}, PhD}
\author[4]{Andrea Herrera \orcidaffil{0000-0003-2898-7161}, PhD}
\author[5]{Sonia Camacho \orcidaffil{0000-0002-3662-1402}, PhD}
\affil[1]{Global Development Institute, University of Manchester, Mánchester, Reino Unido}
\affil[2]{Kalangala Infrastructure Services Ltd., Kalangala, Uganda}
\affil[3]{Departamento de Ingeniería de Sistemas, Pontificia Universidad Javeriana, Bogotá, Colombia}
\affil[4]{Departamento de Ingeniería de Sistemas y Computación, Universidad de los Andes, Bogotá, Colombia}
\affil[5]{Facultad de Administración, Universidad de los Andes, Bogotá, Colombia}

\abstract{La pandemia del VIH/sida es un problema mundial que ha afectado de manera desigual a varios países. Debido a la complejidad de esta condición y al drama humano que representa para aquellos afectados por ésta, múltiples disciplinas han contribuido a solucionar o al menos aliviar esta situación, y el campo de los sistemas de información (SI) no ha estado ausente de estos esfuerzos. Partiendo de la importancia de la terapia antirretroviral (TAR), varias iniciativas en el campo de los SI se han centrado en formas de mejorar la adherencia y la eficacia de esta terapia: los recordatorios por teléfono móvil (para la toma de pastillas y las citas), y las interfaces móviles entre pacientes y personal médico son contribuciones populares. Sin embargo, muchas de estas soluciones han sido difíciles de implantar o desplegar en algunos países del Sur Global, que se encuentran entre los más afectados por esta pandemia. Este artículo presenta uno de estos casos. Utilizando un enfoque de estudio de casos con una técnica de selección de casos extremos, el artículo estudia un sistema de salud móvil (salud-m) para pacientes con VIH en la región de Kalangala (Uganda). Utilizando el modelo de brecha entre diseño-realidad de Heeks para el análisis de datos, el artículo muestra que la rica interacción entre el contexto social y la tecnología debe considerarse como una preocupación central a la hora de diseñar o desplegar tales sistemas.}
\usepackage[citestyle=authoryear,bibstyle=authoryear,backend=biber,maxbibnames=3,isbn=false,uniquename=minfull] {biblatex}
\AtEveryBibitem{\clearfield{pagetotal}}
\date{\today}
\title{Sistemas de información de salud en contextos extremos: Uso de teléfonos móviles para combatir el sida en Uganda}
\hypersetup{
 pdfauthor={Livingstone Njuba, Juan E. Gómez-Morantes, Andrea Herrera, Sonia Camacho},
 pdftitle={Sistemas de información de salud en contextos extremos: Uso de teléfonos móviles para combatir el sida en Uganda},
 pdfkeywords={},
 pdfsubject={},
 pdfcreator={Emacs 29.1 (Org mode 9.7-pre)}, 
 pdflang={Spanish}}
\usepackage{biblatex}
\begin{document}

\maketitle

\TPJcopyright{Este es un artículo de acceso abierto bajo los términos de la Licencia Creative Commons Atribución-NoComercial, la cual permite el uso, distribución y reproducción en cualquier medio, siempre y cuando el trabajo original sea citado correctamente y no se utilice con fines comerciales. Puede encontrar más información sobre la licencia en el siguiente enlace: \href{https://creativecommons.org/licenses/by-nc/4.0/}{CC-BY-NC License}.}

\vspace{.5cm}
\noindent\fbox{\begin{minipage}{\textwidth}
El presente documento es una traducción al español (generada con inteligencia artificial y revisada por los autores) del artículo:
\newline\newline
Njuba, L., Gómez-Morantes, J. E., Herrera, A., \& Camacho, S. (2024). Health information systems in extreme contexts: Using mobile phones to fight AIDS in Uganda. \emph{The Electronic Journal of Information Systems in Developing Countries}, e12314. \url{https://doi.org/10.1002/isd2.12314}
\newline\newline
Por favor usar el artículo original para cualquier cita o referencia.
\newpage
\end{minipage}}

\section{Introducción}
\label{sec-intro}
La pandemia de VIH/sida sigue siendo un problema mundial crítico hasta la fecha. Alrededor de 38,4 millones de personas en todo el mundo vivían con el VIH a finales de 2021, la mayoría de ellas en países de renta baja y media. La situación es especialmente crítica en el África subsahariana, donde viven dos tercios (67\%) de las personas seropositivas. En 2021 se produjeron más de 1,5 millones de nuevas infecciones, lo que representa el menor descenso anual de nuevas infecciones por el VIH desde 2016. Un total de 51\% de estas nuevas infecciones se produjeron en el África subsahariana, en donde mujeres y niñas representaron el 63\% de todas las nuevas infecciones. Además, el África subsahariana representó el 59\% de las 650.000 muertes relacionadas con el sida en 2021 \autocite{UNADIS2022}. Para empeorar las cosas, las crisis que han sacudido el mundo en los últimos dos años han tenido un impacto devastador en las personas que viven con el VIH. Como consecuencia, la respuesta al sida se ha visto sometida a tensiones, y las comunidades que ya corrían un mayor riesgo de contraer el VIH son ahora aún más vulnerables, como el África subsahariana, que sigue siendo la región más afectada por el VIH. Sin embargo, en la región se siguen haciendo esfuerzos para hacer frente a esta pandemia en términos de sensibilización, prevención, pruebas, atención, y tratamiento, y se espera que muchos países de la región alcancen los objetivos 95-95-95\footnote{Más del 95\% de las personas que viven con VIH conocen su estado serológico, más del 95\% de las personas que conocen su estado serológico acceden al tratamiento, y más del 95\% de las personas en tratamiento tienen una carga viral suprimida.}  en breve \autocite{UNADIS2022}.

La importancia de alcanzar estos objetivos radica en que la terapia antirretroviral (TAR) ha transformado esta enfermedad en una afección de largo plazo \autocite{Cooper2017}, resaltando así el valor de la adherencia al tratamiento. Esta última se entiende como el grado en que los pacientes toman la medicación tal y como les ha sido prescrita por sus profesionales sanitarios \autocites{Armitage2020}[][]{Osterberg2005}. Sin embargo, la adherencia al tratamiento sigue siendo un reto importante debido a factores como la mala comunicación entre el paciente y el profesional sanitario \autocites{Haskard2009}[][]{Ondenge2017}, y el olvido \autocites{Dowshen2012}[][]{Kalichman2017}. Se han estudiado diferentes intervenciones para aumentar la adherencia al tratamiento \autocites{Spaan2020}[][]{Whiteley2021}. Aunque ninguna puede mejorar la adherencia por sí sola, se ha sugerido que las tecnologías móviles (por ejemplo, llamadas telefónicas, servicio de mensajes cortos (SMS)) tienen el potencial de mejorar la adherencia al tratamiento en entornos con recursos limitados, como las economías en desarrollo y en transición \autocites{Chib2012}[][]{Cooper2017}[][]{Hirsch-Moverman2017}[][]{Lester2010}[][]{Orr2015}[][]{Siedner2012}. Esas tecnologías ofrecen varias características, como ubicuidad, movilidad, y flexibilidad, que las convierten en una alternativa adecuada para aumentar la adherencia de los pacientes al TAR.

Los resultados de las investigaciones que utilizan tecnologías móviles indican resultados positivos al aumentar la adherencia entre los pacientes con VIH cuando se utilizan este tipo de intervenciones. Sin embargo, también se ha identificado la necesidad de analizar las intervenciones con tecnologías móviles (en adelante, intervenciones de salud-m) entre comunidades específicas \autocites{Catalani2013}[][]{Cooper2017}[][]{Devi2015}[][]{Tufts2015}. Para abordar este vacío, la siguiente pregunta de investigación guía este estudio: ¿Cuáles son los desafíos técnicos, organizativos, y culturales de las intervenciones de salud-m basadas en recordatorios móviles sobre la adherencia al tratamiento del VIH en el Sur Global? Para explorar esta cuestión, se estudian las comunidades pesqueras móviles de Kalangala (Uganda), que presentan características contextuales únicas. El estudio se realiza siguiendo una metodología de estudio de caso único \autocite{Yin2009} utilizando un método de selección de casos \guillemotleft{}extremos\guillemotright{} \autocite{Seawright2008}.

El resto de este artículo se desarrolla del siguiente modo. La sección \ref{sec-revision} describe brevemente la investigación previa sobre tecnologías móviles en la adherencia al tratamiento del VIH. La sección \ref{sec-pregunta} presenta la pregunta de investigación de este estudio, mientras que la sección \ref{sec-metodologia} se centra en la metodología, presentando el contexto de la investigación y los antecedentes teóricos. La sección \ref{sec-hallazgos} presenta los principales hallazgos de este estudio de caso extremo, y la sección \ref{sec-discusion} concluye con un análisis de las principales implicaciones y limitaciones de esta investigación.

\section{Enfoques de salud-m para la adherencia a la TAR}
\label{sec-revision}
La adherencia a los tratamientos puede definirse como el grado en que los pacientes toman los medicamentos según lo prescrito por sus profesionales sanitarios \autocites{Armitage2020}[][]{Osterberg2005} y puede analizarse utilizando puntos de vista orientados al proceso o a los resultados. Una visión orientada a los resultados utiliza el resultado del tratamiento (por ejemplo, la tasa de curación) para medir el éxito, mientras que una visión orientada al proceso emplea variables intermedias (por ejemplo, el recuento de pastillas) para medir la adherencia \autocite{De2003}. Este estudio adopta una visión de la adherencia orientada al proceso porque una intervención digital por sí sola no puede garantizar un resultado concreto del tratamiento, por lo que se descarta un enfoque orientado a los resultados.

Como ya se ha mencionado, el TAR ha transformado el sida en una enfermedad de largo plazo \autocite{Cooper2017}. Sin embargo, la adherencia al tratamiento en pacientes con VIH sigue siendo un desafío debido a varios factores como la mala comunicación entre los proveedores de atención médica y los pacientes, y el olvido \autocites{Dowshen2012}[][]{Martin2005}. Una mala adherencia puede acarrear consecuencias negativas, como la resistencia a los fármacos, una supresión viral menos eficaz, un mayor riesgo de transmisión del VIH, y una mala calidad de vida \autocites{Schneider2004}[][]{Shah2019}.

Pueden introducirse varias intervenciones para mejorar la adherencia al tratamiento. Las intervenciones presenciales requieren invertir recursos y tiempo, y puede resultar difícil para los pacientes acceder a ellas \autocite{Cooper2017}. Por lo tanto, las estrategias, herramientas, y servicios de cibersalud ofrecen una alternativa accesible. En particular, el desarrollo de dispositivos móviles ha permitido intervenciones de salud-m que buscan tratar condiciones de salud como el sida \autocite{Nhavoto2017}. Estas intervenciones pueden proporcionar servicios consistentes a una amplia población a un bajo coste \autocite{Muessig2015}. Se han estudiado diferentes tipos de intervenciones de salud-m para la adherencia al TAR (por ejemplo, llamadas telefónicas, uso de SMS, y aplicaciones), que se describen brevemente a continuación.

La mayoría de las intervenciones se realizan a través de SMS. Estas intervenciones pueden consistir en mensajes unidireccionales ---principalmente recordatorios--- o bidireccionales, en los que los participantes pueden responder confirmando la adherencia o accediendo a más información \autocite{Cooper2017}. Los mensajes tienden a ser encubiertos, sin referirse directamente ni al TAR ni al VIH \autocites{Moore2015}[][]{Sabin2015}. Investigaciones anteriores han analizado los SMS en entornos con recursos limitados como Kenia, Sudáfrica y Uganda \autocites{Crankshaw2010}[][]{Lester2010}[][]{Mitchell2011a}[][]{Pop-Eleches2011}[][]{Rana2015}, destacando su bajo coste y accesibilidad en zonas remotas \autocite{Saberi2011}. En cuanto a los resultados, se ha demostrado que los recordatorios por SMS reducen significativamente las tasas de inasistencia a las citas, aumentan la adherencia a la medicación, y mejoran medidas fisiológicas como el recuento de CD4\footnote{Esta es una prueba para medir la cantidad de células CD4 (un tipo de glóbulo blanco) en la sangre.} o la carga viral\footnote{Esta es una prueba para medir la cantidad de copias de VIH en la sangre. Un TAR exitoso debería reducir la carga viral en el paciente.} \autocite{Mayer2017}. Las intervenciones administradas por SMS son aceptables para los pacientes \autocites{Cele2019}[][]{Cooper2017}, con beneficios reportados tales como ser recordado acerca de las citas o la ingesta de medicamentos, así como un acceso más fácil al apoyo sanitario \autocites{Costa2012}[][]{Smillie2014}. Sin embargo, se deben considerar barreras técnicas como la pérdida de teléfonos o cargadores y la desconexión temporal del servicio \autocite{Norton2014}. Además, deben abordarse preocupaciones sobre la privacidad y la seguridad (por ejemplo, el intercambio involuntario de información sanitaria), especialmente teniendo en cuenta que compartir dispositivos móviles es común en los países africanos \autocites{Cele2019}[][]{Saberi2011}.

Las intervenciones de salud móvil también emplean aplicaciones móviles. Estas aplicaciones permiten a los participantes registrar o hacer un seguimiento de su ingesta de medicación y acceder a información sobre el VIH y el TAR mediante diferentes recursos, como texto, vídeo o juegos \autocites{Horvath2019}[][]{Whiteley2018}. Además, las intervenciones con este tipo de aplicaciones podrían tener la misma finalidad que las intervenciones con SMS descritas anteriormente (por ejemplo, enviar mensajes a los participantes), pero empleando otras aplicaciones populares, como WhatsApp o Facebook (por ejemplo, \cite{Stankievich2018}). Se ha demostrado que este tipo de intervenciones reducen la carga viral y mejoran la adherencia autodeclarada, así como síntomas, como la ansiedad o la neuropatía \autocite{Schnall2018}. Las características interactivas de las aplicaciones también aumentan el compromiso de los participantes y su concienciación sobre el VIH, y la necesidad de cumplir con el TAR \autocite{Cooper2017}. Las aplicaciones móviles pueden ser aceptadas por los pacientes, especialmente si son fáciles de usar, apoyan a los participantes, y no tienen fallas técnicas \autocite{Horvath2019}.

Por último, algunas intervenciones emplean llamadas telefónicas. Investigaciones anteriores indican que las llamadas diarias ayudan a aumentar la adherencia autodeclarada y reducen la carga viral \autocite{Belzer2015}, a diferencia de las llamadas menos frecuentes (por ejemplo, dos llamadas semanales) que no tienen impacto en dichas variables \autocite{Huang2013}. Las intervenciones realizadas a través de llamadas telefónicas suscitan comentarios positivos de los participantes respecto a sentirse apoyados por un facilitador y experimentar una mayor motivación para cumplir con el TAR \autocite{Cooper2017}.

Esta breve revisión indica que las investigaciones anteriores han encontrado resultados positivos de las intervenciones de salud-m para aumentar la adherencia al TAR entre los pacientes con VIH. Sin embargo, las revisiones sobre el tema también destacaron la necesidad de analizar las intervenciones de salud-m en comunidades específicas \autocites{Catalani2013}[][]{Cooper2017}[][]{Devi2015}[][]{Tufts2015}. Es fundamental tener en cuenta las características específicas de dichas comunidades (por ejemplo, la necesidad de adaptar el lenguaje a los diferentes grupos étnicos o el uso de texto en zonas rurales frente a aplicaciones en entornos urbanos) a la hora de explorar la eficacia de dichas intervenciones \autocite{Cele2019}. Por lo tanto, este estudio sigue un caso extremo en Uganda para ampliar el conocimiento actual sobre este tipo de intervenciones y los factores que deben tenerse en cuenta al diseñar dichos sistemas.

\section{Pregunta de investigación}
\label{sec-pregunta}
Esta investigación se centra en la eficacia de las intervenciones de salud electrónica (salud-e), especialmente los recordatorios por teléfono móvil (m-health), para mejorar la adherencia al tratamiento en pacientes con VIH en el Sur Global. Además, la investigación se interesa por los factores (por ejemplo, tecnológicos, sociales, culturales) que median (es decir, mejoran o dificultan) esa eficacia. Dado que la e-salud en el Sur Global es una empresa desafiante \autocites{Agarwal2015}[][]{Braa2004}, es esencial explorar qué factores entran en juego en este tipo de proyectos, de dónde proceden los retos, y cómo pueden conceptualizarse y modelarse.

La cuestión de la conceptualización de los retos ha sido ampliamente debatida por autores como \textcite{Heeks2006}. Sin embargo, todavía hay preguntas abiertas sobre de dónde vienen los desafíos, cómo se manifiestan, y cómo se desarrollan con el tiempo \autocite{Masiero2016a}. Además, estos retos son contingentes, localizados, y están estrechamente vinculados a contextos y condiciones locales \autocite{Heeks2002}. Por lo tanto, cada caso planteará retos diferentes que podrían influir en la eficacia del sistema. Asimismo, cada caso es una nueva oportunidad para aprender algo nuevo sobre los retos de la cibersalud en el Sur Global.

Además, las características inherentes del Sur Global exigen más investigación sobre los retos que pueden surgir en estos contextos, especialmente en regiones o comunidades que se desvían de los casos típicos. Sólo entonces, la comunidad académica podrá desarrollar conceptos, marcos, y estrategias para abordar estos retos, al menos parcialmente. Teniendo esto en cuenta, la pregunta de investigación de este artículo es la siguiente:

PdI: \emph{¿Cuáles son los retos técnicos, organizativos, y culturales de las intervenciones de salud-m basadas en recordatorios móviles sobre la adherencia al tratamiento del VIH en el Sur Global?}

\section{Metodología}
\label{sec-metodologia}
Siguiendo a \textcite[p. 60]{Blaikie2000}, la que se propone en esta investigación es una mezcla entre una pregunta sobre el qué y una pregunta sobre el por qué, con objetivos relacionados con la descripción y la explicación. Además, el hecho de que la cibersalud en general, y no sólo en el Sur Global, sea un fenómeno social que no se puede controlar (descartando así las metodologías basadas en la experimentación) ni predecir (descartando así las metodologías con una fuerte epistemología positivista) lleva a los investigadores a seleccionar un método de estudio de caso. A esta decisión deben seguir otras tres consideraciones: el tipo de diseño de estudio de caso a seguir, la técnica de selección de casos y el diseño y protocolo de la investigación.

Siguiendo la clasificación propuesta por \textcite{Yin2009} y teniendo en cuenta los recursos disponibles para este proyecto, esta investigación se limita a un diseño holístico de caso único. En consonancia con las motivaciones de la pregunta de investigación, este caso único se eligió utilizando un método de selección de casos \guillemotleft{}extremo\guillemotright{}\footnote{Consulte a \textcite{Seawright2008} para ver una discusión detallada sobre otros métodos de selección de casos y una comparación entre ellos.}. Según \textcite[, p. 301]{Seawright2008}, \guillemotleft{}el método del caso extremo selecciona un caso por su valor extremo en la variable independiente (X) o dependiente (Y) de interés\guillemotright{}. La razón de ser de este método de selección es identificar casos altamente inusuales que podrían permitir el descubrimiento de variables independientes previamente ignoradas o ilustrar interacciones previamente no estudiadas entre variables conocidas \autocite{Seawright2016}. Además, un caso extremo podría contribuir más a la literatura actual que un caso en un contexto promedio a la hora de debatir los retos particulares del Sur Global.

A continuación, se presentará el caso (sección \ref{sec-contexto-caso}), seguida del protocolo de investigación (sección \ref{sec-protocolo}). Por último, en la sección \ref{sec-marco-conceptual} se presentará el marco conceptual utilizado para el análisis de los datos.

\subsection{Contexto del caso}
\label{sec-contexto-caso}
El distrito de Kalangala está situado a orillas del lago Victoria, en el suroeste de Uganda. Tiene una superficie de 9066,8 km2, pero sólo 432,1 km2 (4,8\%) es tierra, mientras que el resto es agua. Limita al norte con el distrito de Mpigi, al este con Mukono, al sur con Tanzania, y al oeste con los distritos de Masaka y Rakai. El distrito de Kalangala consta de 84 islas dispersas en el lago Victoria, pero sólo 64 tienen asentamientos humanos. La isla más grande es Bugala, con una superficie de 296 km2. Comprende los condados de Bujumba y Kyamuswa, y siete subcondados de Mugoye, Bujumba, Kalangala Town Council, Kyamuswa, Mazinga, Bufumira, y Bubeke. Para acceder a Kalangala desde el continente, al no haber puentes que unan las islas con el continente, sólo se pueden utilizar dos medios de transporte: por agua (barco/ferry) y por aire. Sólo la isla de Bugala está conectada con el continente por ferry a Masaka y a Entebbe por barco. Además, el transporte entre las islas es complicado y caro. Por ejemplo, se tarda una media de 4,5 horas en viajar de la isla de Bugala a la de Nkose/Lujaabwa (si el lago no tiene vientos desastrosos). Al resto de las islas sólo se puede llegar en embarcaciones locales privadas o públicas, por lo que las islas Kalangala son sobre todo remotas e inaccesibles \autocites{Bwette2014}[][]{Humanitaria2012}.

La mayor parte de la población se dedica a la pesca agrupada a lo largo de las orillas del lago Victoria. La industria pesquera realiza una importante contribución al desarrollo económico nacional, estimada en el 2,3\% del Producto Interior Bruto Nacional \autocite{Statistics2022}. También contribuye a la seguridad alimentaria y a los ingresos familiares. La pesca se complementa con la agricultura, y el aceite de palma se cultiva principalmente en los subcondados de Bujumba y Mugoye, y en el Ayuntamiento de Kalangala \autocite{Monitor2021}.

La última proyección de población para Uganda indica que Kalangala tiene una población total de 69.500 habitantes, con una proporción de mujeres por cada hombre de 1:3 \autocite{Statistics2022}. Esta proporción podría explicar por qué las autoridades han atribuido la propagación del VIH/sida a los hombres que comparten parejas sexuales \autocite{Athman2019}. Además, la mayoría reside en zonas rurales o pesqueras aisladas, lo que dificulta el transporte y el acceso a los servicios clínicos de TAR. Además, en la región se hablan varios idiomas debido a las numerosas tribus que hay en Uganda \autocite{Statistics2022}, y Kalangala es una composición de casi todas las etnias junto con trabajadores migrantes/pescadores de los países vecinos del Congo, Tanzania, Kenia, y Ruanda. Los pescadores de estas comunidades son muy móviles, siempre en busca de mejores caladeros. Del mismo modo, las trabajadoras sexuales suelen desplazarse constantemente en busca de nuevos mercados. Los niveles de alfabetización varían entre las islas, pero la población pesquera de Kalangala es en su mayoría semianalfabeta \autocites{Bwette2014}[][]{Humanitaria2012}.

En cuanto a las capacidades de comunicación, la red de telefonía móvil GSM es bastante buena en el Ayuntamiento de Kalangala. Sin embargo, la red telefónica suele ser deficiente fuera de este municipio, especialmente en las demás islas pequeñas y remotas. Algunos operadores de telecomunicaciones\footnote{Airtel Uganda, Uganda Telecom, y K2 Telecom.} tienen una conectividad de red muy limitada y poco fiable, mientras que otros\footnote{Africell Uganda, Vodafone Uganda, Smile Telecom, y Smart Telecom.} no tienen ninguna. Este problema está relacionado con los elevados costos de instalación de torres de infraestructura de redes de telecomunicaciones por parte de los operadores de telecomunicaciones, que se ven obstaculizados por el terreno geográfico y la espesa vegetación que bloquean la conectividad. Además, aunque la posesión y el uso de teléfonos móviles en Kalangala han ido creciendo con los años, muchos pacientes no tienen números de contacto válidos. Algunos pacientes pierden sus teléfonos y no sustituyen las tarjetas SIM antiguas, sino que compran otras nuevas porque les sale más barato que sustituir la tarjeta SIM perdida.

Además, la propiedad compartida de teléfonos móviles es habitual (por ejemplo, marido y mujer), ya que no todo el mundo puede permitirse un teléfono móvil personal \autocite{Burrell2010}. En adición, de las 64 islas con asentamientos, sólo dos (Buggala y Kitobo) tienen un suministro eléctrico adecuado, lo que dificulta aún más el uso de teléfonos móviles. La gente depende de la energía solar en el resto de las islas para cargar sus teléfonos, mientras que los pocos que pueden permitírselo utilizan generadores. Así, los usuarios de teléfonos móviles pierden a menudo llamadas mientras el teléfono se está cargando.

Los avances en la lucha contra el VIH y el sida entre 2010 y 2020 en Uganda son dignos de elogio, ya que se convirtió en uno de los ocho países del mundo en alcanzar el objetivo 90-90-90 para 2020 \autocite{Commission2021}, garantizando que el 90\% de las personas que viven con el VIH/sida sean conscientes de su estado seropositivo, que el 90\% de los seropositivos reciban tratamiento, y que el 90\% de ellos estén viralmente suprimidos. A pesar de estos logros, la epidemia sigue estando muy extendida, y el distrito de Kalangala sigue siendo el más afectado, con una tasa de prevalencia del 18\%, frente a la tasa nacional del 5,4\% \autocite{Commission2021}. A pesar de las mayores tasas de prevalencia del VIH, las islas de Kalangala sólo cuentan con 15 centros de salud y sólo un centro de salud de nivel IV (KHCIV) actúa como centro de salud regional \autocite{Kwiringira2021}. Además, sólo se organizan nueve clínicas de terapia antirretrovírica en estos centros de salud, normalmente los martes y los jueves. Dado que los centros de salud existentes se encuentran en sólo siete\footnote{Buggala, Bufumira, Lulamba, Bukasa, Bubeke, Mazinga, y Lujjaabwa.} de las 65 islas habitadas, los pacientes deben desplazarse de una isla a otra para buscar el servicio de una clínica de TAR, lo que resulta muy costoso \autocite{Monitor2021a}.

En las clínicas de tratamiento antirretroviral, los historiales médicos de los pacientes con VIH se registran primero en papel y luego en el sistema de historiales médicos abiertos OpenMRS\footnote{OpenMRS es un software de código abierto para gestionar registros médicos. Surgió de una colaboración entre la Escuela de Medicina de la Universidad de Indiana (EE. UU.) y la Universidad Moi (Kenia). Lo administra OpenMRS Inc., una organización sin fines de lucro con sede en Estados Unidos. Recibe apoyo de organizaciones como los Centros para el Control y la Prevención de Enfermedades de Estados Unidos (CDC), el Centro Internacional de Investigaciones para el Desarrollo de Canadá (IDRC), el Instituto Nacional de Salud, el Centro Internacional Fogarty, el Proyecto de Aldeas del Milenio del Instituto de la Tierra (Universidad de Columbia), la Fundación Rockefeller, y la Organización Mundial de la Salud. Consulte \url{https://openmrs.org/}.} y DHIS2\footnote{Un sistema de información de gestión de la salud basado en web de código abierto. Consulte \url{https://dhis2.org/about/}.}. La adherencia al tratamiento se mide por las visitas constantes a la clínica de TAR para recoger las dosis posteriores, acudiendo a las citas, y tomando correctamente el número de pastillas requerido a intervalos específicos (recuento de pastillas en el momento de la siguiente visita). Sólo en la clínica de TAR del KHCIV, cuando se introdujo el OpenMRS, había una gran sala llena de expedientes en papel de pacientes con VIH que habían desaparecido del cuidado\footnote{Un paciente perdido o desaparecido del cuidado es aquel que no volvió a tener contacto con el personal médico}, con sólo un 23\% de pacientes activos en el programa, lo que significa que el 77\% restante se han perdido de la atención, han muerto o han emigrado. Estas cifras son preocupantes y un alto indicador de una adherencia subóptima.

En resumen, el distrito de Kalangala se caracteriza por su elevada tasa de prevalencia del VIH, los escasos medios de transporte a los centros sanitarios y al continente, la cobertura desigual y poco fiable de la red GSM, la limitada cobertura eléctrica, la constante inmigración y emigración de personas y los bajos niveles de alfabetización. Estos factores lo convierten en un contexto rico y complejo para estudiar el uso de intervenciones de salud-m en la lucha contra el sida.

\subsection{Protocolo de investigación}
\label{sec-protocolo}
Esta investigación siguió un enfoque cualitativo y utilizó entrevistas semiestructuradas (por teléfono y en persona) en las que participaron una amplia gama de partes interesadas en el caso. Se prefirieron las entrevistas en persona, pero la lejanía y los problemas logísticos de la región obligaron a los investigadores a realizar entrevistas telefónicas en algunos casos, especialmente con consejeros y equipos sanitarios de aldea (VHT) de islas distantes como Nkose y Kachungwa en el subcondado de Mazinga, y Nkese y Jaana en el subcondado de Bubeke. Las entrevistas telefónicas, aunque subóptimas, permitieron a los investigadores comprobar y experimentar los problemas de cobertura y fiabilidad de la red en la zona. Además, se incluyeron en la recopilación de datos la observación exhaustiva y el análisis de documentos.

Se realizaron un total de 30 entrevistas semiestructuradas. Se seleccionó a los entrevistados de forma intencionada para obtener una amplia cobertura de los tipos de partes implicadas en el caso y su zona geográfica. Las entrevistas se grabaron cuando el entrevistado dio su consentimiento explícito. Los informantes clave elegidos en esta investigación fueron clínicos de TAR, responsables de adherencia, consejeros de clínicas de TAR, representantes de ONG de apoyo al VIH/sida, VHT, y partidarios del tratamiento. Nunca se entrevistó a los pacientes por razones éticas y de vulnerabilidad. Sin embargo, las partes interesadas entrevistadas tienen amplia experiencia y poseen información que prácticamente nadie más tiene, y son autoridades en sus áreas. Además, debido a su estrecha relación y constante interacción con los pacientes, son la representación más cercana de las opiniones de los pacientes que puede incluirse en esta investigación sin incumplimientos éticos.

Las lenguas utilizadas para las entrevistas fueron principalmente el inglés y el luganda, según la elección del entrevistado. Cuando los entrevistados no dominaban estas lenguas, se recurrió a la ayuda de un intérprete. Todas las entrevistas que no se realizaron en inglés se tradujeron al inglés durante el proceso de transcripción.

\subsection{Análisis de datos y marco conceptual}
\label{sec-marco-conceptual}
\begin{figure}[htbp]
\centering
\includegraphics[width=.9\linewidth]{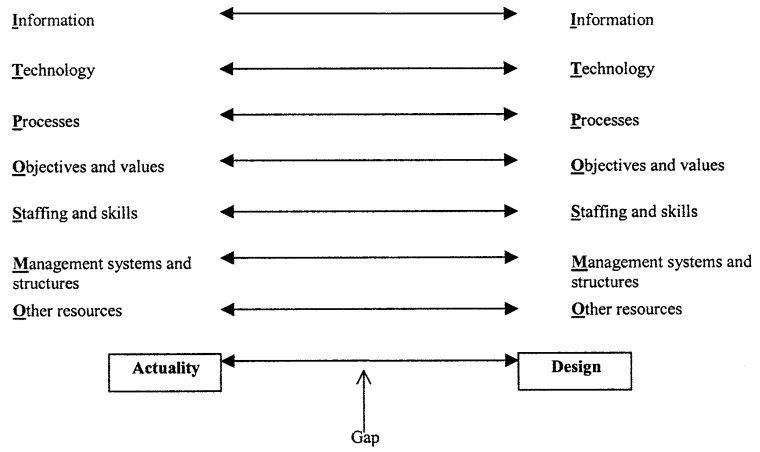}
\caption{\label{fig-drgap}Brechas entre diseño y realidad. Fuente: \autocite{Heeks2002}.}
\end{figure}

Como se menciona en la sección \ref{sec-contexto-caso}, este estudio de casos ha seguido una técnica de selección de casos extremos. Como tal, el marco conceptual utilizado para analizar estos datos debe proporcionar herramientas adecuadas para comprender las circunstancias de los casos extremos o no habituales. Basándose en estos requisitos, los investigadores seleccionan el modelo de brecha diseño-realidad propuesto por Heeks \autocite*{Heeks2002,Heeks2002a} como marco conceptual principal de esta investigación. Este modelo trata de comprender las brechas entre el diseño y la realidad de una determinada iniciativa TIC o basada en TIC \autocite{Heeks2002a}: los aspectos que los diseñadores de las TIC asumen o dan por sentados, pero que no se corresponden con la realidad local en la que se utilizará la TIC. Estas brechas pueden surgir de muchas formas o dimensiones, como muestra el modelo ITPOSMO\footnote{Este es un acrónimo en inglés para Information, Technology, Processes, Objectives and values, Staffing and skills, Management systems and structures, y Other resources.} en la Figura \ref{fig-drgap}.

Las brechas son el resultado de la distancia física o contextual entre las partes interesadas dominantes en el proceso de diseño de una TIC concreta y el contexto en el que dicha TIC se desplegará y utilizará. Un ejemplo clásico es el de los profesionales de las TI en el Norte Global, que basan el diseño de las TIC en la disponibilidad generalizada de conexión a Internet de banda ancha, alfabetización (concretamente en inglés) o cobertura de energía eléctrica que pueden no ser la norma en el Sur Global. Sin embargo, los agentes locales también pueden ser una fuente de brechas entre diseño y la realidad. Esto puede ocurrir, por ejemplo, cuando los agentes locales se educan en instituciones extranjeras (situadas principalmente en el Norte Global) y actúan como caballos de Troya \autocite[, p. 108]{Heeks2002a} para diseños o heurísticas de diseño incoherentes o incompatibles con las realidades locales.

La importancia de comprender estas brechas de diseño-realidad radica en que las brechas podrían provocar el fracaso\footnote{En esta discusión adoptamos la definición de éxito, fracaso parcial, y fracaso propuesta por \textcite{Heeks2002}. En esta definición, una iniciativa exitosa es aquella en la que la mayoría de las partes interesadas logran sus objetivos principales y no experimentan resultados adversos significativos. Un fracaso parcial es una iniciativa que no puede alcanzar los objetivos principales, o una en la que se alcanzaron esos objetivos pero con resultados adversos significativos. Finalmente, una iniciativa fallida es aquella que nunca se implementa o una que se implementa pero se abandona inmediatamente.} de la TIC o de la iniciativa basada en las TIC \autocite{Heeks2002a}. En otras palabras, la probabilidad de éxito o fracaso de una iniciativa basada en las TIC (especialmente en el Sur Global) puede entenderse en función del tamaño de las brechas. Esto es especialmente importante, ya que algunas brechas pueden crear diferencias irreconciliables entre una TIC y su contexto. Por ejemplo, una TIC diseñada bajo el supuesto de una amplia cobertura de energía eléctrica fracasará si esta cobertura no existe. Del mismo modo, una iniciativa diseñada bajo el supuesto de que los datos gubernamentales son fiables, están actualizados y son de fácil acceso fracasará si tan solo uno de estos supuestos está ausente en el contexto local. Por lo tanto, las brechas entre el diseño y la realidad deben comprenderse y gestionarse a lo largo del ciclo de vida de la iniciativa para reducir las posibilidades de fracaso.

Por último, se elige este modelo porque tiene una buena trayectoria en la investigación sobre las TIC en los países en desarrollo \autocites{Bati2021}[][]{Gwamba2018}[][]{Hawari2010}[][]{Kubuga2021}[][]{Masiero2016a}.

\section{Hallazgos}
\label{sec-hallazgos}
El OpenMRS se introdujo en 2016 en todos los centros de salud con clínicas de TAR en el distrito de Kalangala como parte del plan de despliegue nacional. OpenMRS es un sistema cliente-servidor, lo que significa que está diseñado para funcionar en un entorno en el que muchos ordenadores cliente acceden a la misma información en un servidor. Los objetivos de la introducción de este sistema eran mejorar el almacenamiento, el procesamiento, y la gestión de los historiales de los pacientes con VIH/sida, que hasta entonces sólo se conservaban en archivos de datos con fines informativos. También se preveía que el sistema mejoraría la adherencia al tratamiento mediante el envío automático de mensajes SMS diarios a los pacientes con recordatorios sobre la toma de medicación, y que generaría registros/listas de citas que podrían utilizarse para llamar a los pacientes y recordarles sus citas en la clínica de TAR.

A continuación se analizan y califican las brechas diseño-realidad entre los requisitos y supuestos del diseño del sistema de recordatorios móviles y la realidad en el contexto de la comunidad pesquera de Kalangala tras la implantación de los recordatorios móviles, según las dimensiones ITPOSMO. Los autores califican cada brecha de 0 a 10 basándose en las pruebas encontradas sobre el terreno y utilizando los criterios resumidos en la Tabla \ref{tab-criterios-calificación}.

\rowcolors{2}{black!15}{white}
\begin{table}[htbp]
\caption{\label{tab-criterios-calificación}Criterios para la calificación de brechas. Basado en \textcite{Heeks2008b}.}
\centering
\begin{tabular}{ll}
\toprule
Rango & Probabilidad de que la dimensión contribuya al fracaso\\[0pt]
\midrule
8.1–10 & Muy probable\\[0pt]
6.1–8.0 & Probable\\[0pt]
4.1–6.0 & Posible\\[0pt]
2.1–4.0 & Poco probable\\[0pt]
0.0–2.0 & Muy poco probable\\[0pt]
\bottomrule
\end{tabular}
\end{table}

\subsection{Información - Calificación: 7.0/10}
\label{sec-informacion}
Las brechas en esta dimensión muestran la diferencia entre los requisitos de información y los supuestos del diseño del sistema de recordatorio móvil y la realidad en las clínicas de tratamiento antirretrovírico de Kalangala tras su implantación. El diseño partía de la base de que todos los pacientes disponían de historiales médicos precisos, completos y actualizados ---incluidos los contactos de sus teléfonos móviles--- y de los contactos telefónicos de sus respectivas personas de apoyo al tratamiento. En realidad, los historiales médicos de muchos pacientes eran inexactos, estaban incompletos y no contenían los contactos telefónicos de los pacientes. Por ejemplo, en el KHCIV, sólo el 20\% de las historias clínicas capturadas inicialmente en el sistema tenían contactos telefónicos.

Cuando los pacientes seropositivos acuden por primera vez a la clínica de tratamiento antirretrovírico, además de facilitar información personal, se les pide el número de teléfono móvil de las personas de las que prefieren recibir apoyo durante su tratamiento y que incluso puedan recoger los medicamentos si es necesario. Sin embargo, como dijo un asesor entrevistado en el KHCIV: \guillemotleft{}En algunos casos, los pacientes no dan información correcta y completa sobre sí mismos; por ejemplo, un paciente puede darnos nombres y contactos telefónicos erróneos. También puede ocurrir que algunos pacientes no tengan realmente un teléfono móvil, pero otros se niegan deliberadamente a facilitar contactos telefónicos [alegando] que no tienen teléfono móvil. Esto ocurre especialmente con pacientes femeninas que comparten teléfonos móviles con sus parejas y que no les han revelado que están en tratamiento contra el VIH. También es frecuente con aquellos pacientes que todavía se sientan estigmatizados y no quieren [exponerse]\guillemotright{}. El orientador añadió que \guillemotleft{}cuando tenemos los números de contacto de las personas que apoyan el tratamiento, acabamos descubriendo estos casos al hablar con ellas (\ldots{}) comparten con nosotros la información correcta del paciente; de lo contrario, es muy difícil ponerse en contacto con este tipo de pacientes\guillemotright{}.

Además, a veces los pacientes dan información incoherente sobre sí mismos en diferentes visitas a la clínica de TAR y acaban con registros múltiples (y contradictorios) en el OpenMRS. Un clínico de TAR informó de que esto ocurre cuando el paciente teme ser juzgado por no acudir a las citas de la clínica de TAR. También se dan casos de pacientes que cambian irregularmente de número de teléfono, especialmente cuando los teléfonos se pierden o se dañan durante la pesca. Teniendo en cuenta que el coste de sustituir la tarjeta SIM es más elevado y debe hacerse desde Kampala, la mayoría de los pacientes prefieren comprar una nueva tarjeta SIM.

\subsection{Tecnología - Calificación: 7.5/10}
\label{sec-tecnologia}
Las brechas en esta dimensión muestran la diferencia entre los requisitos tecnológicos y los supuestos del diseño del sistema de recordatorios móviles y la realidad en las clínicas de TAR y en la comunidad pesquera de Kalangala tras la implantación de los recordatorios móviles. El diseño asumía la disponibilidad de ordenadores personales adecuados para ejecutar el sistema OpenMRS en todas las clínicas de TAR, con suministro eléctrico fiable y señal de red móvil para enviar mensajes SMS diarios a los pacientes. Además, el diseño suponía que todos los pacientes adultos disponían de teléfonos móviles con buena cobertura GSM en el lugar donde vivían, capacidad para cargar sus teléfonos y datos móviles suficiente para recibir los SMS.

La clínica KHCIV TAR dispone de tres ordenadores con el sistema OpenMRS, mientras que el resto tiene un ordenador cada una. Sin embargo, un funcionario de EMR\footnote{Los oficiales EMR gestionan la información de los pacientes (es decir, captura, actualización, generación de informes) en OpenMRS y DHIS2.} señaló que \guillemotleft{}estos ordenadores fallan a menudo como consecuencia del desgaste normal o del mal uso por parte de los usuarios, y cuando los ordenadores de las islas más alejadas, como Mazinga, se averían, puede pasar incluso un mes antes de que se reparen porque hay que llevarlos a la isla principal, Bugala, para que los reparen.Es costoso alquilar un barco privado para que un técnico informático vaya a una isla como ésta a reparar un solo ordenador\guillemotright{}. En cuanto a la conectividad a Internet, las clínicas TAR utilizan \emph{dongles} de Internet móvil para conectarse y suscribirse a datos de Internet cada trimestre. Sin embargo, algunos centros tienen una conectividad a internet poco fiable y lenta debido a la escasa cobertura de la red de telecomunicaciones, y otros se quedan sin datos antes de que acabe el trimestre, dejando el sistema sin internet. Este problema se puso de manifiesto en el Centro de Salud III (HCIII) de Bukasa, donde los datos rara vez duran más de un mes. En cuanto a la electricidad, de las 84 islas, sólo la mayor (Bugala) dispone de electricidad. El resto de las islas utilizan energía solar, inversores o generadores para hacer funcionar los sistemas informáticos de los centros de salud y las clínicas de terapia antirretrovírica. En cuanto a la carga de los teléfonos, la mayoría de los pacientes de las islas aisladas suelen cargarlos en lugares públicos. Durante un grupo focal, algunos de los que apoyan el tratamiento en la isla de Bubeke destacaron que \guillemotleft{}solemos cargar los teléfonos en videotecas o cines públicos porque tienen energía solar o generadores, pero los teléfonos suelen tardar mucho en cargarse por completo, y los generadores y la energía solar dañan las baterías de nuestros teléfonos\guillemotright{} Un miembro de VHT de la misma isla también planteó la preocupación de que \guillemotleft{}mientras se cargan los teléfonos en esos lugares públicos es posible que llegue el recordatorio del SMS, y otras personas pueden leer el SMS en el teléfono de los pacientes, lo que va en contra de la confidencialidad\guillemotright{}.

La posesión de teléfonos móviles está aumentando. Sin embargo, muchos pacientes siguen sin tener teléfonos móviles personales. El director de la clínica de tratamiento antirretrovírico del KHCIV calcula que el 40\% de los pacientes no tiene teléfono móvil personal, y del 60\% restante, sólo el 20\% tiene smartphone. En otras palabras, la propiedad compartida de teléfonos móviles entre los isleños es habitual. Como afirmó un consejero de la clínica TAR de KHCIV: \guillemotleft{}El porcentaje de pacientes que comparten teléfonos con sus parejas es de alrededor del 30\%, y uno se da cuenta de que un buen número de ellos no han revelado su estado serológico a sus parejas, por lo que uno puede llamar a un paciente y si su pareja responde a la llamada, puede ser complicado\guillemotright{}. Un VHT de la zona pesquera de Mwena informó de que una de las razones por las que se comparten los teléfonos móviles entre las parejas es que algunos hombres se niegan deliberadamente a que sus esposas tengan teléfonos móviles por miedo a la infidelidad. Además, los pacientes suelen estar ilocalizables cuando sus teléfonos no están cargados o sin señal. En cuanto a los datos móviles , en la mayoría de las islas remotas, el coste de un bono de datos móviles se incrementa en 100 o 200 UGX en comparación con el coste en el continente. Por último, la cobertura de la red telefónica es escasa en las islas. Un educador sanitario del Proyecto Integral de Servicios de Salud Pública de Kalangala (KCPHSP), que trabaja en colaboración con las clínicas de terapia antirretrovírica y visita periódicamente las islas de Kalangala para educar a los pacientes con VIH y proporcionarles material doméstico, declaró que \guillemotleft{}la mayoría de las islas tienen una cobertura de red muy deficiente para la mayoría de los operadores de telecomunicaciones. En algunas islas, como Lwabaswa, Bubembe, y Kisujju, los pacientes deben ir a puntos específicos de las islas para poder hacer llamadas con el operador Airtel, que es muy popular porque sus tarifas de llamadas son bajas.\guillemotright{} Un partidario del tratamiento de la isla de Bubembe planteó una preocupación similar y añadió que \guillemotleft{}si recibimos una llamada y no estamos en esos puntos concretos de la red, no estamos disponibles, así que, para hacer y recibir llamadas, nos aseguramos de desplazarnos a esos puntos de la red\guillemotright{}. Uno de los investigadores señaló también que incluso el operador de telecomunicaciones con mejor cobertura de red de Uganda (MTN) no cubre toda la isla de Bugala y tiene algunos puntos oscuros desde los que no se pueden hacer ni recibir llamadas.

\subsection{Proceso - Calificación: 8.5/10}
\label{sec-procesos}
Las brechas en esta dimensión muestran la diferencia entre las prácticas y procesos organizativos previstos al diseñar el sistema de recordatorio móvil y la realidad en las clínicas de TAR tras su implantación. El diseño partía de la base de que, al acudir por primera vez a la clínica de TAR tras dar positivo, el paciente recibe el asesoramiento adecuado y se le aconseja que inicie el tratamiento. El historial médico y la información personal de los pacientes se registran en un archivo en papel cuando aceptan el tratamiento. A continuación, el encargado de la gestión de los historiales médicos los introduce en OpenMRS. En visitas posteriores, se comprueba si falta algún dato en el historial del paciente y se actualiza en consecuencia. A continuación, el OpenMRS procesa la información del paciente y genera recordatorios por SMS que se envían automáticamente a los pacientes en función de las fechas de las próximas citas y los horarios de toma de medicación. Además, el sistema genera una lista de pacientes cuya próxima cita está programada en las próximas 48 h para que el responsable de adherencia o el clínico responsable llame a los pacientes y les recuerde las citas. También se asumió en el diseño que los registros de los pacientes se capturan en tiempo real o inmediatamente después de la visita a la clínica de TAR para que los pacientes empiecen a recibir recordatorios móviles lo antes posible y se les llame y recuerde de forma proactiva las próximas citas de la clínica de TAR antes de la fecha prevista.

En realidad, los recordatorios telefónicos se suelen hacer después de que los pacientes falten a una cita en la clínica de TARV y no de forma proactiva, como supone el diseño. El responsable de la adherencia informó de que los pacientes que faltan a tres citas consecutivas con el centro de TAR se clasifican como perdidos para el seguimiento o perdidos para la atención y como abandonados si faltan a una cuarta cita. Por otro lado, un asesor dijo: \guillemotleft{}A veces damos toda la información al responsable de adherencia cuando los pacientes tienen que acudir a las próximas citas de la clínica TAR, pero a veces se olvida de llamar a los pacientes y acaba llevando los expedientes en su coche, donde incluso se extravían.\guillemotright{} Además, los historiales médicos no se registran en tiempo real. Por ejemplo, hasta hace poco, el KHCIV seguía capturando datos atrasados de pacientes almacenados en carpetas de archivos desde 2004, cuando empezó el tratamiento del VIH en Kalangala. \guillemotleft{}Durante el primer año de implantación de OpenMRS, sólo teníamos registros de más de 8.000 pacientes, por lo que el proceso de captura de la información ha llevado más tiempo del esperado\guillemotright{}, dijo el responsable de TAR del KHCIV. Y añadió: \guillemotleft{}De momento, los nuevos pacientes tardan algún tiempo en recibir recordatorios móviles\guillemotright{}.

Además, los historiales de los pacientes en las clínicas de terapia antirretrovírica rara vez se actualizan de forma continua. El oficial de ERM anotó que \guillemotleft{}algunos consejeros de clínicas de TAR muestran reticencia a actualizar los registros de pacientes cuando estos regresan para visitas posteriores; por ejemplo, muchos de los expedientes de pacientes que no proporcionaron números de teléfono en la visita inicial en las clínicas de TAR en KHCIV y Mugoye han permanecido sin esos números de contacto debido a que la mayoría de los consejeros tienden a ignorar el proceso normal de actualizar la información faltante de los pacientes en visitas posteriores.\guillemotright{}

\subsection{Objetivos y valores - Calificación: 8.5/10}
\label{sec-objetivos}
Las brechas en esta dimensión muestran las brechas entre los objetivos y valores (es decir, a través de los cuales se manifiestan la cultura y la política) del diseño del sistema de recordatorio móvil y la realidad tras su implantación. El diseño del sistema partía de la base de que se lograría un objetivo de reducción de costes en comparación con las visitas físicas a los pacientes en sus domicilios. Además, se fijaron otros objetivos, como el seguimiento de los pacientes perdidos, el aumento de la adherencia, y la mejora de las decisiones de gestión. Sin embargo, algunas partes interesadas no compartían algunos de esos objetivos y valores. Les preocupaba la confidencialidad y privacidad de los pacientes, que podrían verse vulneradas con los recordatorios móviles. Otras partes interesadas, como el responsable de adherencia y los médicos de las distintas clínicas de TAR, consideraban que aprender el nuevo sistema era estresante. También les resultaba pesado utilizar OpenMRS, que genera las citas de los pacientes y la lista de contactos para los recordatorios móviles. Esta resistencia se debió principalmente a unos conocimientos informáticos moderados, al miedo a perder el empleo, y al aumento de la carga de trabajo, según informó la persona de contacto del EMR de Uganda en el distrito de Kalangala. Además, aunque un escenario favorable podría haber supuesto una reducción de costes para el sistema sanitario, también habría supuesto un aumento de los costes para los pacientes, que ahora tendrán que hacer frente a costes más elevados asociados al funcionamiento de sus teléfonos móviles, como más recargas de datos móviles, tasas de sustitución de la tarjeta SIM, sustitución rápida de teléfonos perdidos o dañados, y mantener el teléfono cargado. Dada la vulnerabilidad socioeconómica de la población general de Kalangala, esto parece una transferencia inaceptable de costes del sistema nacional de salud a los pacientes.

El diseño también partía de la base de que no existen restricciones culturales en cuanto a la propiedad y el uso de teléfonos móviles personales entre los adultos. Los valores culturales de la comunidad pesquera local de Kalangala no impiden a ningún adulto poseer teléfonos móviles. Sin embargo, algunos hombres de las comunidades pesqueras de Kalangala disuaden a sus esposas de poseer teléfonos móviles personales. En tales situaciones, asocian el uso del teléfono móvil con la ruptura familiar.

En cuanto a los recordatorios móviles, en el diseño del sistema se asumió que serían cómodos para los pacientes y que la privacidad no sería un problema; por ello, se contactó con los pacientes o se les enviaron recordatorios por teléfono. De hecho, los recordatorios móviles son prácticos para una pequeña parte de los pacientes que han declarado públicamente que viven con el VIH/sida. Para la mayoría de los pacientes que no lo han dado a conocer a los demás, la privacidad es una grave preocupación, agravada por la propiedad compartida de los dispositivos móviles. Temían el estigma dirigido por la comunidad. El responsable de adherencia señaló que \guillemotleft{}a los pacientes que lo han declarado públicamente no les preocupa la privacidad cuando reciben estos recordatorios móviles y hablan cómodamente sobre el VIH/sida con total libertad; sin embargo, los que no lo han hecho, ni siquiera quieren que vaya a sus casas porque, por lo general, el público piensa que cada vez que voy a una determinada casa, hay un paciente con VIH que no está tomando bien la medicación\guillemotright{}. En la misma línea, el médico encargado de la terapia antirretrovírica en el KHCIV advirtió de que la confidencialidad dependía de cómo se enfrentara el paciente al estigma. Los pacientes que han revelado públicamente que viven con el VIH y tienen una buena adherencia, aunque sean pocos, se denominan \guillemotleft{}pacientes/clientes expertos\guillemotright{}. Esos pacientes expertos desempeñan un papel fundamental a la hora de ayudar a otros a enfrentarse al estigma y a revelar su estado para mejorar su adherencia. Con la revelación continua, la cuestión de la privacidad y el estigma se volverá menos crítica.

\subsection{Dotación de personal y competencias - Calificación: 6.0/10}
\label{sec-personal}
Las brechas en esta dimensión muestran la diferencia entre las competencias cualitativas y cuantitativas esperadas al diseñar el sistema de recordatorio móvil y la realidad de las clínicas de TAR y los pacientes después de la implementación. El diseño partía de la base de que el personal de las clínicas de terapia antirretrovírica (para utilizar el OpenMRS) y los pacientes (para utilizar los teléfonos móviles) debían poseer un amplio abanico de competencias y conocimientos. Por ejemplo, se asumió que todo el personal de las clínicas de TAR tendría conocimientos sobre OpenMRS o sistemas similares. También se consideró que todos los pacientes pueden utilizar teléfonos móviles (por ejemplo, hacer llamadas telefónicas, enviar y leer mensajes de texto).

En cuanto al personal de las clínicas de terapia antirretrovírica, varios asesores y clínicos de terapia antirretrovírica recibieron formación especializada para utilizar el sistema OpenMRS. Tras la formación, formaron a otros. Sin embargo, el encargado de la adherencia no estaba familiarizado con el sistema, por lo que se basaba en los datos de los asesores y los encargados del EMR para realizar las llamadas. El clínico responsable de TAR señaló que \guillemotleft{}aunque muchos asesores/clínicos de TAR y algunos clientes expertos con conocimientos informáticos han recibido formación sobre el uso del sistema OpenMRS, la velocidad de introducción de datos sigue siendo baja, ya que los usuarios no tienen mucha experiencia y, por razones similares, el sistema no se utiliza de forma óptima\guillemotright{}. Los niveles de dotación de personal se consideran suficientemente buenos, ya que hay cuatro asesores, un responsable de adherencia, un cliente experto y clínicos de TAR que ayudan a introducir los datos en el OpenMRS.

En cuanto a los pacientes, los conocimientos para utilizar el dispositivo móvil en la comunidad de pescadores no eran los esperados. Debido a los altos niveles de analfabetismo, muchos pacientes no pueden enviar ni leer mensajes de texto en sus teléfonos. En tales casos, se limitan a ignorar los SMS. En el peor de los casos, los pacientes llevan sus teléfonos móviles a un tercero para que lea los mensajes por ellos, lo que supone un claro problema de privacidad.

\subsection{Sistemas y estructuras de gestión - Calificación: 4.0/10}
\label{sec-gestion}
Esta dimensión representa las brechas entre los sistemas y estructuras de gestión necesarios para la implantación y el funcionamiento satisfactorios del sistema móvil de recordatorios OpenMRS y la gestión y estructuras reales en los centros de salud tras la implantación. El diseño requería una gestión y una toma de decisiones descentralizadas, que permitieran que las decisiones se tomaran a nivel de departamento de la clínica de TAR. Además, para que el sistema tuviera éxito se necesitaba un considerable apoyo de la alta dirección.

De hecho, la descentralización era considerable y las decisiones se tomaban a nivel departamental. El clínico de TAR del KHCIV señaló: \guillemotleft{}Podemos organizar por nuestra cuenta campañas comunitarias de promoción de la adherencia y enviar recordatorios móviles a los pacientes, y sólo tenemos que avisar a la dirección. La alta dirección nos apoya mucho en lo que respecta a los recordatorios móviles, y da datos móviles al responsable de adherencia, al clínico responsable de TAR, y al responsable de PTMI\footnote{Prevención de transmisión de madre a infante.} para llamar a los pacientes en un intento de mejorar la adherencia\guillemotright{}. Aunque la alta dirección se mostró muy favorable, a veces se vio desbordada por las peticiones de facilitación.

\subsection{Otros recursos: Tiempo y dinero - Calificación: 5.0/10}
\label{sec-otros}
Esta dimensión representa las brechas entre los recursos necesarios (por ejemplo, dinero y tiempo) para la implementación y el funcionamiento satisfactorios del diseño de OpenMRS y los recordatorios móviles y la realidad de los recursos disponibles en las clínicas de TAR y la comunidad de pescadores de Kalangala. El diseño partía de la base de que el personal de las clínicas de terapia antirretrovírica siempre dispondría de datos móviles para llamar a los pacientes y recordarles sus citas en las clínicas de terapia antirretrovírica y la medicación programada. El KCPHSP proporcionó trimestralmente el dinero para realizar las llamadas telefónicas y enviar los SMS.

Por lo que respecta a los pacientes, cargar los teléfonos móviles y comprar datos móviles en algunas islas resultaba caro. Por ejemplo, un promotor del tratamiento de la isla de Lwabaswa señaló que \guillemotleft{}normalmente pagamos entre 500 y 1.000 UGX por cargar un teléfono móvil y, debido a las baterías defectuosas, puede que haya que cargarlo tres veces en una semana. En nuestra isla ha aumentado el costo de las llamada y se cobran precios diferentes; un bono de 500 UGX en el continente cuesta aquí 600 UGX, y uno de 1.000 UGX cuesta 1.200 UGX, y podemos pasar incluso tres días sin un solo bono de datos móviles en todas las tiendas de aquí antes de que lo traigan de la ciudad de Kalangala\guillemotright{}.

La situación se agravó por la campaña \guillemotleft{}\emph{Stop Illegal Fishing}\guillemotright{} administrada por el ejército ugandés en el lago Victoria, que dejó a muchos pacientes de Kalangala sin trabajo y más pobres tras quemar por la fuerza sus aparejos de pesca y detener a algunos. Muchos pescadores ni siquiera podían permitirse una comida regular que les permitiera tomar los medicamentos. Esto afectó a la capacidad de los pacientes para comprar datos móviles y cargar sus teléfonos; simplemente luchaban por ganarse la vida. Por este motivo, muchos pacientes de la isla de Kaagoonya habían sido clasificados como pacientes perdidos al cuidado. Cuando recibieron la visita de los médicos especialistas en terapia antirretrovírica, contaron que habían vendido sus teléfonos móviles para sobrevivir y que ni siquiera habían podido reunir el dinero del transporte para ir a las clínicas de terapia antirretrovírica a recoger los medicamentos. Puede decirse que, aunque el sistema pudiera enviar el recordatorio por móvil en condiciones óptimas, habría sido inútil, ya que algunos pacientes ni siquiera podían recoger la medicación. Y aunque así fuera, algunos de ellos ni siquiera podían permitirse la comida necesaria para tomar la medicación.

\subsection{Evaluación global del sistema basada en la calificación de las deficiencias}
\label{sec-evaluacion-global}
Tabla \ref{tab-resumen-calificaciones} presenta un resumen de las calificaciones de las deficiencias para cada una de las dimensiones de la ITPOSMO y la calificación agregada de las deficiencias de esta iniciativa:

\rowcolors{2}{black!15}{white}
\begin{table}[htbp]
\caption{\label{tab-resumen-calificaciones}Resumen de las calificaciones para las calificaciones ITPOSMO}
\centering
\begin{tabular}{ll}
\toprule
Dimensión & Calificación\\[0pt]
\midrule
Información & 7.0\\[0pt]
Tecnología & 7.5\\[0pt]
Procesos & 8.5\\[0pt]
Objetivos y valores & 8.5\\[0pt]
Personal y competencias & 6.0\\[0pt]
Sistemas y estructuras de gestión & 4.0\\[0pt]
Otros recursos & 5.0\\[0pt]
\midrule
Calificación total & 46.5/70\\[0pt]
\bottomrule
\end{tabular}
\end{table}

Siguiendo la escala propuesta por \textcite{Heeks2003}, esta iniciativa tiene una calificación alta y es probable que fracase a menos que se tomen medidas concretas para zanjar algunas brechas. Debe prestarse especial atención a las dimensiones de tecnología, procesos, y objetivos y valores. La dimensión de la información también tiene una calificación crítica, pero la mayoría de los problemas de esa dimensión se deben a una recogida de datos inadecuada en los procesos asociados a la iniciativa o a valores específicos de la región que impiden una recogida de datos suficiente.

\section{Discusión y conclusiones}
\label{sec-discusion}
Esta sección comienza resumiendo los principales retos encontrados en el caso. Después, se retoma la pregunta de investigación presentada en la sección \ref{sec-pregunta} para explicar cómo se respondió a ella. A continuación, se analizan las conclusiones presentadas hasta ahora en el artículo desde dos perspectivas: las implicaciones para la bibliografía general sobre sistemas de información en el Sur Global y para los profesionales de este campo, y las contribuciones a la bibliografía sobre adherencia al tratamiento antirretroviral y salud móvil. Por último, se discuten las limitaciones del artículo.

\subsection{Resumen de casos prácticos}
\label{sec-resumen}
La Tabla \ref{tab-resumen-desafios} muestra un resumen de los retos detectados en el estudio de caso. También clasifica los retos por categorías (técnicos, organizativos, culturales) y dimensiones ITPOSMO (información, tecnología, procesos, objetivos y valores, personal y competencias, sistemas y estructuras de gestión, otros recursos).

\rowcolors{2}{white}{white}
\begin{longtable}{p{.2\textwidth}p{.2\textwidth}p{.5\textwidth}}
\caption{\label{tab-resumen-desafios}Resumen de los desafíos encontrados en el caso de estudio}
\\[0pt]
\toprule
Categoría del desafío & Dimensión & Desafío\\[0pt]
\midrule
\endfirsthead
\multicolumn{3}{l}{Continúa de la página anterior} \\[0pt]
\toprule

Categoría del desafío & Dimensión & Desafío \\[0pt]

\midrule
\endhead
\midrule\multicolumn{3}{r}{Continúa en la siguiente página} \\
\endfoot
\endlastfoot
Técnico & Información & Pérdida de dispositivos móviles que conduce al cambio de números de teléfono de los pacientes.\\[0pt]
 &  & Múltiples registros con diferentes teléfonos móviles.\\[0pt]
 & Tecnología & Fallos en los ordenadores\\[0pt]
 &  & Altos costos de reparación\\[0pt]
 &  & Conectividad a internet poco confiable y lenta\\[0pt]
 &  & Generadores y cargas solares dañando las baterías de los teléfonos de los pacientes.\\[0pt]
 &  & Alto costo de recargas de datos móviles en islas remotas\\[0pt]
 &  & Cobertura de red en áreas remotas\\[0pt]
 & Procesos & \\[0pt]
 & Objetivos y valores & Mayores costos para los pacientes (por ejemplo, datos móviles, tarifas de reemplazo de SIM)\\[0pt]
 & Personal y competencias & \\[0pt]
 & Sistemas y estructuras de gestión & \\[0pt]
 & Otros recursos & Alto costo asociado a la carga de teléfonos móviles de los pacientes\\[0pt]
\cellcolor{black!15}Organizacional & \cellcolor{black!15}Información & \cellcolor{black!15}Registros médicos inexactos e incompletos de los pacientes.\\[0pt]
\cellcolor{black!15} & \cellcolor{black!15}Tecnología & \cellcolor{black!15}\\[0pt]
\cellcolor{black!15} & \cellcolor{black!15}Procesos & \cellcolor{black!15}Recordatorios por llamada telefónica solo se hacen después de una cita perdida.\\[0pt]
\cellcolor{black!15} & \cellcolor{black!15} & \cellcolor{black!15}Registros médicos físicos de los pacientes llevados fuera de las clínicas y potencialmente extraviados.\\[0pt]
\cellcolor{black!15} & \cellcolor{black!15} & \cellcolor{black!15}Captura de datos retrasada\\[0pt]
\cellcolor{black!15} & \cellcolor{black!15} & \cellcolor{black!15}Registros de pacientes no actualizados en visitas posteriores.\\[0pt]
\cellcolor{black!15} & \cellcolor{black!15}Objetivos y valores & \cellcolor{black!15}Aprender a usar el sistema OpenMRS se considera estresante y su uso como una carga.\\[0pt]
\cellcolor{black!15} & \cellcolor{black!15}Personal y competencias & \cellcolor{black!15}No todo el personal está capacitado y familiarizado con el sistema\\[0pt]
\cellcolor{black!15} & \cellcolor{black!15}Sistemas y estructuras de gestión & \cellcolor{black!15}La alta dirección a veces se ve abrumada con solicitudes de facilitación.\\[0pt]
\cellcolor{black!15} & \cellcolor{black!15}Otros recursos & \cellcolor{black!15}\\[0pt]
Cultural & Información & Los pacientes dan información errónea de contacto o se niegan a proporcionar información\\[0pt]
 & Tecnología & Cargar los teléfonos de los pacientes en lugares públicos conduce a posibles violaciones de la privacidad ya que otros pueden leer los mensajes de texto.\\[0pt]
 &  & Compartir dispositivos móviles con parejas\\[0pt]
 &  & Reducción de la propiedad femenina de dispositivos, controlados por sus maridos.\\[0pt]
 & Procesos & \\[0pt]
 & Objetivos y valores & La propiedad femenina de teléfonos móviles asociada con rupturas familiares.\\[0pt]
 &  & Preocupaciones de privacidad, debido a no divulgar el estado de VIH a otros.\\[0pt]
 &  & Percepción negativa de las visitas de los oficiales de adherencia.\\[0pt]
 & Personal y competencias & Niveles de analfabetismo entre los pacientes que les impide leer o enviar mensajes de texto.\\[0pt]
 & Sistemas y estructuras de gestión & \\[0pt]
 & Otros recursos & Restricciones económicas de los pacientes para poder comprar medicamentos.\\[0pt]
\bottomrule
\end{longtable}

\subsection{Retos técnicos, organizativos y culturales de las intervenciones de salud-m en el Sur Global}
\label{sec-retos}
La principal pregunta de investigación de este artículo es: ¿Cuáles son los retos técnicos, organizativos y culturales de las intervenciones de salud-m basadas en recordatorios móviles sobre la adherencia al tratamiento del VIH en el Sur Global? Esta pregunta, sin embargo, no puede responderse de forma generalizable (véase la sección \ref{sec-implicaciones-tar}); hacerlo sería contradecir la importancia de un diseño contextualizado que se ha comentado anteriormente en este artículo. A su vez, este artículo responde a esta pregunta utilizando un método de casos extremos para identificar y debatir retos técnicos, organizativos, y culturales inusuales que afectan a la intervención de salud-m en el Sur Global (véase la sección \ref{sec-hallazgos}). Aunque estos retos surgen de un caso particular y extremo y distan mucho de ser casos medios, es precisamente este carácter único lo que los convierte en una contribución significativa; se trata de retos raramente tratados en la bibliografía relativa a las intervenciones de salud-m.

\subsection{Implicaciones para la literatura general sobre sistemas de información en el Sur Global}
\label{sec-implicaciones-si}
Este artículo utiliza el modelo de brecha diseño-realidad propuesto por \textcite{Heeks2002}. Para ello, aporta un nuevo estudio de caso que valida la capacidad explicativa del modelo. En efecto, el modelo fue capaz de captar una amplia gama de elementos de diseño y heurísticos integrados en las soluciones tradicionales de las TIC del Norte Global que chocan con los contextos locales del Sur Global. Como muestra la literatura, este choque podría poner en peligro la eficacia o la sostenibilidad de los sistemas de información en el Sur Global. Los profesionales podrían utilizar este conocimiento para ser más críticos con las suposiciones que se hacen al transferir tecnología, diseños, o heurísticas de diseño de una región a otra. Para ello, el modelo de brecha diseño-realidad podría ser útil como herramienta de evaluación previa para contrastar los diseños con los contextos locales y mejorarlos antes de los esfuerzos de implantación, elevando así el diseño al contexto local \autocite{Heeks2006}. Del mismo modo, este análisis puede revelar cómo es necesario elevar el contexto al diseño. En esos casos, es esencial incluir estas necesidades como parte integrante del proyecto o iniciativa.

Una segunda implicación es la necesidad de utilizar una perspectiva sociotécnica \autocite{Avgerou1998} para entender los sistemas de información como sistemas sociales que hacen un uso intensivo de información. Esto es importante en cualquier contexto, pero especialmente en el Sur Global: como demuestra el caso de Kalangala, los sistemas sociales del Sur Global pueden ser extremadamente diferentes de los del Norte Global. Estas diferencias no pueden simplemente ignorarse y deben ser una preocupación central de los esfuerzos de diseño. En consecuencia, la lógica utilizada para abordar estos sistemas en el Sur Global debe ser diferente y adaptarse a los contextos locales. El uso de lo que \textcite[, p. 110]{Heeks2002} denomina \guillemotleft{}híbridos\guillemotright{} (personas que entienden el contexto, la organización, los procesos y cómo estos elementos se relacionan con los sistemas de información) o campeones de proyecto \autocite{Renken2019} es un consejo habitual.

Asumir el ahorro de costes o el aumento de la eficiencia como valores centrales en los proyectos de sistemas de información es un buen ejemplo de la cuestión mencionada. Dado que la mayor parte de los costes ahorrados proceden de la sustitución de mano de obra humana por tecnología automatizada, otros valores, como la seguridad en el empleo, se ven afectados. Dado que la pérdida de un puesto de trabajo puede ser (mucho) más difícil de recuperar en el Sur Global, las implicaciones éticas y prácticas de esta decisión son mucho más complejas en este entorno; esta vulnerabilidad debería tener mayor prioridad que los valores corporativos o empresariales. Además, la idea de que la tecnología automatizada es más barata que la mano de obra humana, aunque suele ser cierta en el Norte Global, no se aplica necesariamente al Sur Global \autocite{Heeks2002a}.

Por último, este caso muestra la importancia de ampliar la concepción del propio proyecto de sistema de información. Una evaluación previa de cualquier sistema de información en el Sur Global que utilice el modelo de la brecha diseño-realidad desvelará retos como los analizados en la sección \ref{sec-hallazgos}. Algunos de estos retos podrían abordarse tecnológicamente adaptando los diseños a las realidades locales. Sin embargo, otros elementos no técnicos del sistema de información necesitan respuestas diferentes que deberían incluirse en el proyecto o iniciativa. Por ejemplo, en un contexto como el de la región de Kalangala, en el que la cobertura de energía eléctrica es escasa y poco fiable, el proyecto podría utilizar un principio de bajo consumo energético para diseñar aplicaciones o interfaces de servicio que consuman la menor cantidad de energía posible (una respuesta a nivel técnico), o el proyecto podría incluir el suministro de cargadores solares portátiles para los usuarios finales del sistema de información (una respuesta a nivel sociotécnico).

\subsection{Contribuciones a la literatura sobre adherencia al TAR y salud-m}
\label{sec-implicaciones-tar}
Este artículo amplía la bibliografía que analiza las mejoras en la adherencia al tratamiento en pacientes con VIH que utilizan salud-m en entornos con recursos limitados. Estudios anteriores realizados en esos entornos han examinado la viabilidad de esas intervenciones, su utilidad como programas de prevención, las mejoras en la adherencia al tratamiento, y las sugerencias de los pacientes para que esas intervenciones sean más eficaces (por ejemplo, Kenia, Sudáfrica, Uganda; \cites{Crankshaw2010}[][]{Lester2010}[][]{Mitchell2011a}[][]{Pop-Eleches2011}[][]{Rana2015}). Aunque esos estudios emplearon diferentes metodologías de investigación (por ejemplo, ensayos controlados aleatorizados, encuestas, grupos focales), siempre se centraron en los pacientes y prestaron menos atención al sistema sanitario (y social) que los rodea. Según el leal saber y entender de los autores, esta investigación es el primer estudio que analiza las brechas en el diseño y la aplicación de intervenciones de salud-m para la adherencia al TAR en un caso extremo teniendo en cuenta las perspectivas de los diferentes actores en la prestación de servicios de salud para pacientes con VIH. Además, el análisis incluye una amplia gama de factores sociales, técnicos y económicos.

Quizá la consecuencia más importante de esta investigación para los investigadores y profesionales de la salud-m sea que las iniciativas de salud-m deben diseñarse desde una perspectiva sociotécnica amplia para ofrecer soluciones holísticas a problemas como la insuficiencia de ingresos, la conectividad, la disponibilidad de energía eléctrica, una dotación de personal escasa o inadecuada, la falta de datos fiables u otros muchos que pueden surgir en diferentes contextos. Para lograrlo, se aconseja una visión sociotécnica de los sistemas de información, de modo que los artefactos técnicos de los sistemas de información (por ejemplo, la aplicación de software) no eclipsen otros elementos cruciales que pueden crear brechas de diseño-realidad lo suficientemente grandes como para poner en peligro la iniciativa.

Siguiendo esta idea, es esencial destacar que los factores que dificultan la eficacia de las intervenciones de salud-m no se limitan al paciente (por ejemplo, teléfono móvil compartido, analfabetismo), sino que también pueden provenir de los proveedores de atención sanitaria (por ejemplo, falta de información completa, dificultades para utilizar los sistemas de historia clínica, personal insuficiente), o del contexto más amplio (por ejemplo, falta de cobertura de la red móvil, red eléctrica poco fiable, vulnerabilidad económica generalizada). Esto pone de relieve la utilidad del modelo de brecha diseño-realidad para abordar de forma holística el diseño de intervenciones de salud-m basadas en SMS.

\subsection{Limitaciones}
\label{sec-limitaciones}
Como cualquier otro proyecto de investigación, éste está sujeto a limitaciones que afectan tanto a los resultados como a su generalizabilidad.

Aunque el modelo de la brecha entre diseño y realidad resultó útil para responder a la pregunta planteada en esta investigación, la comprensión de cómo se formaron estas brechas y, sobre todo, cómo resolverlas queda fuera del alcance del modelo; en él no hay ninguna herramienta teórica o metodológica explícita para abordar estas cuestiones\footnote{Cabe destacar que \textcite{Heeks2003} ha discutido estrategias genéricas para la reducción de brechas en el contexto de proyectos de eGovernment, pero estas no son un componente central del modelo de brecha entre el diseño y la realidad.}. Esta limitación del modelo de brecha de diseño-realidad ha sido señalada por otras investigadoras como \textcite{Masiero2016a}, quien ha realizado importantes contribuciones para ampliar el modelo. Esto, sin embargo, tiene poco impacto en esta investigación porque su pregunta de investigación está alineada con las fortalezas del modelo de brecha de diseño-realidad (es decir, identificación y clasificación de factores y desafíos que afectan la efectividad de las intervenciones digitales en el Sur Global).

Una de las principales limitaciones metodológicas de esta investigación fue no incluir a pacientes reales en el proceso de investigación. Esta decisión se tomó por motivos éticos para proteger la intimidad de las personas vulnerables. Mientras que los investigadores de otras regiones podían incluir a pacientes con VIH en sus investigaciones sin ponerlos en peligro, las condiciones de la región de Kalangala y el estigma social que allí se atribuye al VIH fueron el principal factor de esta exclusión. No obstante, la perspectiva de los pacientes se incluyó indirectamente a través de los distintos agentes sanitarios que interactúan con ellos a diario. Los investigadores confían en que esta exclusión, aunque subóptima, no tuvo un impacto significativo en los resultados de la investigación.

Por último, está la cuestión de la generalizabilidad de los resultados. En coherencia con la visión sociotécnica de los sistemas de información expuesta en la sección \ref{sec-implicaciones-si}, esta investigación no pretende ser generalizable de forma tradicional (relacionando variables independientes con variables dependientes sin importar el contexto). Esto se debe a que, desde esta perspectiva, los sistemas de información son sistemas sociales con un uso intensivo de la información, y los sistemas sociales son, por naturaleza, particulares y están limitados por el contexto. Por último, la selección de un caso extremo hace más difícil encontrar un caso lo suficientemente similar como para que esto sea incluso factible. Sin embargo, esto no reduce el valor de esta investigación, ya que proporciona conocimientos prácticos y lecciones que pueden generalizarse en forma de principios de diseño o elementos teóricos que son tan importantes como la generalización tradicional \autocites{Flyvbjerg2006}[][]{Walsham1995}.

\printbibliography

@article{Seawright2008,
  title =	 {Case {{Selection Techniques}} in {{Case Study
                  Research}}: {{A Menu}} of {{Qualitative}} and
                  {{Quantitative Options}}},
  author =	 {Seawright, Jason and Gerring, John},
  date =	 2008,
  journaltitle = {Political Research Quarterly},
  shortjournal = {Political Research Quarterly},
  volume =	 61,
  number =	 2,
  pages =	 {294--308},
  issn =	 {1065-9129, 1938-274X},
  doi =		 {10.1177/1065912907313077}
}

@article{Heeks2002,
  title =	 {Information {{Systems}} and {{Developing
                  Countries}}: {{Failure}}, {{Success}}, and {{Local
                  Improvisations}}},
  author =	 {Heeks, Richard},
  date =	 2002,
  journaltitle = {The Information Society},
  shortjournal = {The Information Society},
  volume =	 18,
  number =	 2,
  pages =	 {101--112},
  doi =		 {10.1080/01972240290075039}
}

@online{Heeks2003,
  title =	 {Most {{eGovernment-for-Development Projects Fail}}:
                  {{How Can Risks}} Be {{Reduced}}?},
  author =	 {Heeks, Richard},
  date =	 2003,
  shortjournal = {SSRN},
  issn =	 {1556-5068},
  doi =		 {10.2139/ssrn.3540052},
}

@article{Walsham1995,
  title =	 {Interpretive {{Case Studies}} in {{Is Research}}:
                  {{Nature}} and {{Method}}},
  author =	 {Walsham, G.},
  date =	 1995,
  journaltitle = {European Journal of Information Systems},
  volume =	 4,
  number =	 2,
  pages =	 {74--81},
  doi =		 {10.1057/ejis.1995.9}
}

@article{Flyvbjerg2006,
  title =	 {Five {{Misunderstandings About Case-Study
                  Research}}},
  author =	 {Flyvbjerg, Bent},
  date =	 2006,
  journaltitle = {Qualitative Inquiry},
  shortjournal = {Qualitative Inquiry},
  volume =	 12,
  number =	 2,
  pages =	 {219--245},
  issn =	 {1077-8004, 1552-7565},
  doi =		 {10.1177/1077800405284363}
}

@article{Masiero2016a,
  title =	 {The Origins of Failure: {{Seeking}} the Causes of
                  {{Design}}–{{Reality}} Gaps},
  author =	 {Masiero, S.},
  date =	 2016,
  journaltitle = {Information Technology for Development},
  volume =	 22,
  number =	 3,
  pages =	 {487--502},
  doi =		 {10.1080/02681102.2016.1143346},
}

@article{Rana2015,
  author =	 {Rana, Yashodhara and Haberer, Jessica and Huang,
                  Haijing and Kambugu, Andrew and Mukasa, Barbara and
                  Thirumurthy, Harsha and Wabukala, Peter and Wagner,
                  Glenn J. and Linnemayr, Sebastian},
  title =	 {Short {Message} {Service} ({SMS})-{Based}
                  {Intervention} to {Improve} {Treatment} {Adherence}
                  among {HIV}-{Positive} {Youth} in {Uganda}: {Focus}
                  {Group} {Findings}},
  journaltitle = {PLOS ONE},
  date =	 2015,
  volume =	 10,
  number =	 4,
  pages =	 {e0125187},
  issn =	 {1932-6203},
  doi =		 {10.1371/journal.pone.0125187},
}

@article{Pop-Eleches2011,
  author =	 {Pop-Eleches, Cristian and Thirumurthy, Harsha and
                  Habyarimana, James P. and Zivin, Joshua G. and
                  Goldstein, Markus P. and de Walque, Damien and
                  MacKeen, Leslie and Haberer, Jessica and Kimaiyo,
                  Sylvester and Sidle, John and Ngare, Duncan and
                  Bangsberg, David R.},
  title =	 {Mobile phone technologies improve adherence to
                  antiretroviral treatment in a resource-limited
                  setting: a randomized controlled trial of text
                  message reminders},
  journaltitle = {AIDS},
  date =	 2011,
  volume =	 25,
  number =	 6,
  pages =	 825,
  issn =	 {0269-9370},
  doi =		 {10.1097/QAD.0b013e32834380c1},
}

@article{Mitchell2011a,
  title =	 {Cell phone usage among adolescents in {Uganda}:
                  acceptability for relaying health information},
  volume =	 26,
  issn =	 {0268-1153, 1465-3648},
  doi =		 {10.1093/her/cyr022},
  number =	 5,
  journaltitle = {Health Education Research},
  author =	 {Mitchell, K. J. and Bull, S. and Kiwanuka, J. and
                  Ybarra, M. L.},
  date =	 2011,
  pages =	 {770--781},
}

@article{Lester2010,
  author =	 {Lester, Richard T and Ritvo, Paul and Mills, Edward
                  J and Kariri, Antony and Karanja, Sarah and Chung,
                  Michael H and Jack, William and Habyarimana, James
                  and Sadatsafavi, Mohsen and Najafzadeh, Mehdi and
                  Marra, Carlo A and Estambale, Benson and Ngugi,
                  Elizabeth and Ball, T Blake and Thabane, Lehana and
                  Gelmon, Lawrence J and Kimani, Joshua and Ackers,
                  Marta and Plummer, Francis A},
  title =	 {Effects of a mobile phone short message service on
                  antiretroviral treatment adherence in Kenya (WelTel
                  Kenya1): a randomised trial},
  journaltitle = {The Lancet},
  date =	 2010,
  volume =	 376,
  number =	 9755,
  pages =	 {1838--1845},
  issn =	 {0140-6736},
  doi =		 {10.1016/s0140-6736(10)61997-6},
}

@article{Crankshaw2010,
  author =	 {Crankshaw, Tamaryn and Corless, Inge B. and Giddy,
                  Janet and Nicholas, Patrice K. and Eichbaum, Quentin
                  and Butler, Lisa M.},
  title =	 {Exploring the Patterns of Use and the Feasibility of
                  Using Cellular Phones for Clinic Appointment
                  Reminders and Adherence Messages in an
                  Antiretroviral Treatment Clinic, Durban, South
                  Africa},
  journaltitle = {AIDS Patient Care and STDs},
  date =	 2010,
  volume =	 24,
  number =	 11,
  pages =	 {729--734},
  issn =	 {1087-2914},
  doi =		 {10.1089/apc.2010.0146},
}

@article{Heeks2002a,
  title =	 {E-Government in Africa: {{Promise}} and Practice},
  author =	 {Heeks, Richard},
  date =	 2002,
  journaltitle = {Information Polity},
  volume =	 7,
  number =	 {2-3},
  pages =	 {97--114},
  doi =		 {10.3233/ip-2002-0008}
}

@article{Renken2019,
  author =	 {Renken, Jaco and Heeks, Richard},
  title =	 {Champions of {IS} {Innovations}},
  journaltitle = {Communications of the Association for Information
                  Systems},
  date =	 2019,
  volume =	 44,
  number =	 1,
  issn =	 {1529-3181},
  doi =		 {10.17705/1CAIS.04438},
}

@inbook{Avgerou1998,
  title =	 {Approaches to Information Systems Development},
  booktitle =	 {Developing {{Information Systems}}},
  author =	 {Avgerou, Chrisanthi and Cornford, Tony},
  date =	 1998,
  pages =	 {161--184},
  publisher =	 {{Macmillan Education UK}},
  location =	 {{London}},
  doi =		 {10.1007/978-1-349-14813-4_8},
  bookauthor =	 {Avgerou, Chrisanthi and Cornford, Tony},
  isbn =	 {978-0-333-73231-1 978-1-349-14813-4},
  langid =	 {english}
}

@article{Heeks2006,
  title =	 {Health Information Systems: {{Failure}}, Success and
                  Improvisation},
  author =	 {Heeks, Richard},
  date =	 2006,
  journaltitle = {International Journal of Medical Informatics},
  volume =	 75,
  number =	 2,
  pages =	 {125--137},
  doi =		 {10.1016/j.ijmedinf.2005.07.024},
}

@article{Kubuga2021,
  author =	 {Kubuga, Kennedy Kumangkem and Ayoung, Daniel
                  Azerikatoa and Bekoe, Stephen},
  title =	 {Ghana’s ICT4AD policy: between policy and reality},
  journaltitle = {Digital Policy, Regulation and Governance},
  date =	 2021,
  volume =	 23,
  number =	 2,
  pages =	 {132--153},
  issn =	 {2398-5038},
  doi =		 {10.1108/DPRG-02-2020-0020},
}

@article{Hawari2010,
  title =	 {Explaining {{ERP}} Failure in a Developing Country:
                  {{A Jordanian}} Case Study},
  author =	 {Hawari, A. and Heeks, R.},
  date =	 2010,
  journaltitle = {Journal of Enterprise Information Management},
  volume =	 23,
  number =	 2,
  pages =	 {135--160},
  doi =		 {10.1108/17410391011019741}
}

@article{Gwamba2018,
  author =	 {Gwamba, G. and Mayende, G. and Isabwe, G.M.N. and
                  Birevu Muyinda, P.},
  title =	 {Conceptualising Design of Learning Management
                  Systems to Address Institutional Realities},
  journaltitle = {Advances in Intelligent Systems and Computing},
  date =	 2018,
  volume =	 716,
  pages =	 {43--50},
  doi =		 {10.1007/978-3-319-73204-6_6}
}

@article{Bati2021,
  title =	 {Evaluating Integrated Use of Information
                  Technologies in Secondary Schools of {{Ethiopia}}
                  Using Design-Reality Gap Analysis: {{A}}
                  School-Level Study},
  author =	 {Bati, Tesfaye Bayu and Workneh, Anteneh Wasyhun},
  date =	 2021,
  journaltitle = {Electronic Journal of Information Systems in
                  Developing Countries},
  volume =	 87,
  number =	 1,
  doi =		 {10.1002/isd2.12148},
}

@online{Monitor2021a,
  title =	 {Long road to good health services in {Kalangala}},
  organization = {Monitor},
  author = {Monitor},
  url =
                  {https://www.monitor.co.ug/uganda/magazines/healthy-living/long-road-to-good-health-services-in-kalangala-1637320},
  urldate =	 {2024-02-26},
  date =	 2021,
}

@article{Kwiringira2021,
  author =	 {Kwiringira, Japheth Nkiriyehe and Mugisha, James and
                  Akugizibwe, Mathias and Ariho, Paulino},
  title =	 {‘When will the doctor be around so that I come by?!’
                  Geo-socio effects on health care supply, access and
                  utilisation: experiences from Kalangala Islands,
                  Uganda},
  journaltitle = {BMC Health Services Research},
  date =	 2021,
  volume =	 21,
  number =	 1,
  pages =	 1163,
  issn =	 {1472-6963},
  doi =		 {10.1186/s12913-021-07204-7},
}

@Online{Commission2021,
  author =	 {{Uganda Aids Commission}},
  title =	 {Facts on HIV and AIDS in Uganda},
  date =	 2021,
  url =
                  {https://uac.go.ug/media/attachments/2021/09/13/final-2021-hiv-aids-factsheet.pdf}
}

@Article{Burrell2010,
  author =	 {Burrell, Jenna},
  title =	 {Evaluating Shared Access: Social equality and the
                  circulation of mobile phones in rural Uganda},
  journaltitle = {Journal of Computer-Mediated Communication},
  date =	 2010,
  volume =	 15,
  number =	 2,
  pages =	 {230--250},
  doi =		 {10.1111/j.1083-6101.2010.01518.x},
}

@online{Humanitaria2012,
  author =	 {{The New Humanitarian}},
  title =	 {{Fishing} communities missing out on {HIV}
                  treatment},
  date =	 2012,
  url =
                  {https://www.thenewhumanitarian.org/news/2012/11/26/fishing-communities-missing-out-hiv-treatment},
  urldate =	 {2024-02-27},
}

@MastersThesis{Bwette2014,
  author =	 {Bwette, Diana},
  title =	 {The Islands of Kalangala District and Access to
                  Antiretroviral Treatment: A Question of Human Rights
                  and Global Health Justice},
  date =	 2014,
  school = 	 {Erasmus University},
  url =		 {https://thesis.eur.nl/pub/17386},
}

@Online{Statistics2022,
  author =	 {{Uganda Bureau of Statistics – UBOS}},
  title =	 {UBOS Statistical Abstract 2021},
  date =	 2022,
  url =
                  {http://library.health.go.ug/publications/statistics/ubos-statistical-abstract-2021},
  urldate =	 {2023-08-16},
}

@Online{Athman2019,
  author =	 {Athman, Raziah},
  title =	 {The Island and HIV: girls fighting for social
                  justice},
  date =	 2019,
  url =
                  {https://www.africanews.com/2019/03/07/the-island-and-hiv-girls-fighting-for-social-justice/},
  urldate =	 {2024-02-26},
  journaltitle = {Africanews},
}

@online{Monitor2021,
  author =	 {{Monitor}},
  title =	 {Kalangala: The 84 islands of fish, oil and vast
                  tourism potential},
  date =	 2021,
  url =
                  {https://www.monitor.co.ug/uganda/special-reports/kalangala-the-84-islands-of-fish-oil-and-vast-tourism-potential-1558918},
  organization = {Monitor},
  urldate =	 {2024-02-26},
}

@Article{Seawright2016,
  author =	 {Seawright, Jason},
  title =	 {The Case for Selecting Cases That Are Deviant or
                  Extreme on the Independent Variable},
  journaltitle = {Sociological Methods \& Research},
  date =	 2016,
  volume =	 45,
  number =	 3,
  pages =	 {493–525},
  issn =	 {1552-8294},
  doi =		 {10.1177/0049124116643556},
}

@book{Yin2009,
  author =	 {Yin, Robert K.},
  title =	 {Case Study Research: Design and Methods},
  date =	 2009,
  edition =	 {4th ed},
  series =	 {Applied Social Research Methods},
  publisher =	 {{Sage Publications}},
  location =	 {{Los Angeles, CA, USA}},
  isbn =	 {978-1-4129-6099-1},
  pagetotal =	 219
}

@book{Blaikie2000,
  title =	 {Designing Social Research: The Logic of
                  Anticipation},
  shorttitle =	 {Designing Social Research},
  author =	 {Blaikie, Norman},
  date =	 2000,
  publisher =	 {{Polity Press}},
  location =	 {{Cambridge, UK}},
  isbn =	 {978-0-7456-1766-4 978-0-7456-1767-1},
  pagetotal =	 338
}

@article{Braa2004,
  author =	 {Braa, Jørn and Monteiro, Eric and Sahay, Sundeep},
  title =	 {Networks of Action: Sustainable Health Information
                  Systems across Developing Countries},
  journaltitle = {MIS Quarterly},
  date =	 2004,
  volume =	 28,
  number =	 3,
  pages =	 {337--362},
  issn =	 {0276-7783},
  doi =		 {10.2307/25148643},
  urldate =	 {2024-02-26},
}

@article{Agarwal2015,
  author =	 {Agarwal, Smisha and Perry, Henry B. and Long,
                  Lesley‐Anne and Labrique, Alain B.},
  title =	 {Evidence on feasibility and effective use of mH
                  ealth strategies by frontline health workers in
                  developing countries: systematic review},
  journaltitle = {Tropical Medicine \& International Health},
  date =	 2015,
  volume =	 20,
  number =	 8,
  pages =	 {1003--1014},
  issn =	 {1360-2276, 1365-3156},
  doi =		 {10.1111/tmi.12525},
}

@article{Cele2019,
  author =	 {Cele, Mthokozisi A. and Archary, Moherndran},
  title =	 {Acceptability of short text messages to support
                  treatment adherence among adolescents living with
                  HIV in a rural and urban clinic in KwaZulu-Natal},
  journaltitle = {Southern African Journal of HIV Medicine},
  date =	 2019,
  volume =	 20,
  number =	 1,
  pages =	 6,
  issn =	 {2078-6751},
  doi =		 {10.4102/sajhivmed.v20i1.976},
}

@article{Tufts2015,
  author =	 {Tufts, Kimberly Adams and Johnson, Kaprea F. and
                  Shepherd, Jewel Goodman and Lee, Ju-Young and Bait
                  Ajzoon, Muna S. and Mahan, Lauren B. and Kim, Miyong
                  T.},
  title =	 {Novel Interventions for HIV Self-management in
                  African American Women: A Systematic Review of
                  mHealth Interventions},
  journaltitle = {Journal of the Association of Nurses in AIDS Care},
  date =	 2015,
  volume =	 26,
  number =	 2,
  pages =	 139,
  issn =	 {1055-3290},
  doi =		 {10.1016/j.jana.2014.08.002},
}

@article{Devi2015,
  author =	 {Devi, Balla Rama and Syed-Abdul, Shabbir and Kumar,
                  Arun and Iqbal, Usman and Nguyen, Phung-Anh and Li,
                  Yu-Chuan (Jack) and Jian, Wen-Shan},
  title =	 {mHealth: An updated systematic review with a focus
                  on HIV/AIDS and tuberculosis long term management
                  using mobile phones},
  journaltitle = {Computer Methods and Programs in Biomedicine},
  date =	 2015,
  volume =	 122,
  number =	 2,
  pages =	 {257--265},
  issn =	 {0169-2607},
  doi =		 {10.1016/j.cmpb.2015.08.003},
}

@article{Cooper2017,
  author =	 {Cooper, Vanessa and Clatworthy, Jane and Whetham,
                  Jennifer and {EmERGE Consortium}},
  title =	 {mHealth Interventions To Support Self-Management In
                  HIV: A Systematic Review},
  journaltitle = {The Open AIDS Journal},
  date =	 2017,
  volume =	 11,
  number =	 1,
  pages =	 {119--132},
  doi =		 {10.2174/1874613601711010119},
}

@article{Catalani2013,
  author =	 {Catalani, Caricia and Philbrick, William and Fraser,
                  Hamish and Mechael, Patricia, and Israelski, Dennis
                  M.},
  title =	 {mHealth for HIV Treatment \& Prevention: A
                  Systematic Review of the Literature},
  journaltitle = {The Open AIDS Journal},
  date =	 2013,
  volume =	 7,
  number =	 1,
  pages =	 {17--41},
  issn =	 {1874-6136},
  doi =		 {10.2174/1874613620130812003},
}

@article{Huang2013,
  author =	 {Huang, Dongsheng and Sangthong, Rassamee and McNeil,
                  Edward and Chongsuvivatwong, Virasakdi and Zheng,
                  Weibin and Yang, Xuemei},
  title =	 {Effects of a Phone Call Intervention to Promote
                  Adherence to Antiretroviral Therapy and Quality of
                  Life of HIV/AIDS Patients in Baoshan, China: A
                  Randomized Controlled Trial},
  journaltitle = {AIDS Research and Treatment},
  date =	 2013,
  volume =	 2013,
  pages =	 {e580974},
  issn =	 {2090-1240},
  doi =		 {10.1155/2013/580974},
}

@article{Belzer2015,
  author =	 {Belzer, Marvin E. and Kolmodin MacDonell, Karen and
                  Clark, Leslie F. and Huang, Jennifer and Olson,
                  Johanna and Kahana, Shoshana Y. and Naar, Sylvie and
                  Sarr, Moussa and Thornton, Sarah},
  title =	 {Acceptability and Feasibility of a Cell Phone
                  Support Intervention for Youth Living with {HIV} with
                  Nonadherence to Antiretroviral Therapy},
  journaltitle = {{AIDS} Patient Care and {STDs}},
  date =	 2015,
  volume =	 29,
  number =	 6,
  pages =	 {338--345},
  issn =	 {1087-2914},
  doi =		 {10.1089/apc.2014.0282},
}

@article{Horvath2019,
  author =	 {Horvath, Keith J. and Lammert, Sara and MacLehose,
                  Richard F. and Danh, Thu and Baker, Jason V. and
                  Carrico, Adam W.},
  title =	 {A Pilot Study of a Mobile App to Support HIV
                  Antiretroviral Therapy Adherence Among Men Who Have
                  Sex with Men Who Use Stimulants},
  journaltitle = {AIDS and Behavior},
  date =	 2019,
  volume =	 23,
  number =	 11,
  pages =	 {3184--3198},
  issn =	 {1573-3254},
  doi =		 {10.1007/s10461-019-02597-3},
}

@article{Schnall2018,
  author =	 {Schnall, Rebecca and Cho, Hwayoung and Mangone,
                  Alexander and Pichon, Adrienne and Jia, Haomiao},
  title =	 {Mobile {Health} {Technology} for {Improving}
                  {Symptom} {Management} in {Low} {Income} {Persons}
                  {Living} with {HIV}},
  journaltitle = {AIDS and Behavior},
  date =	 2018,
  volume =	 22,
  number =	 10,
  pages =	 {3373--3383},
  issn =	 {1573-3254},
  doi =		 {10.1007/s10461-017-2014-0},
}

@article{Stankievich2018,
  author =	 {Stankievich, Erica and Malanca, Adriana and
                  Foradori, Irene and Ivalo, Silvina and Losso,
                  Marcelo},
  title =	 {Utility of {Mobile} {Communication} {Devices} as a
                  {Tool} to {Improve} {Adherence} to {Antiretroviral}
                  {Treatment} in {HIV}-infected {Children} and {Young}
                  {Adults} in {Argentina}},
  journaltitle = {The Pediatric Infectious Disease Journal},
  date =	 2018,
  volume =	 37,
  number =	 4,
  pages =	 345,
  issn =	 {0891-3668},
  doi =		 {10.1097/INF.0000000000001807},
}

@article{Whiteley2018,
  author =	 {Whiteley, Laura and Brown, Larry K. and Mena,
                  Leandro and Craker, Lacey and Arnold, Trisha},
  title =	 {Enhancing health among youth living with {HIV} using
                  an {iPhone} game},
  journaltitle = {AIDS Care},
  date =	 2018,
  volume =	 30,
  number =	 {sup4},
  pages =	 {21--33},
  issn =	 {0954-0121, 1360-0451},
  doi =		 {10.1080/09540121.2018.1503224},
}

@article{Saberi2011,
  author =	 {Saberi, Parya and Johnson, Mallory O.},
  title =	 {Technology-{Based} {Self}-{Care} {Methods} of
                  {Improving} {Antiretroviral} {Adherence}: {A}
                  {Systematic} {Review}},
  journaltitle = {PLOS ONE},
  date =	 2011,
  volume =	 6,
  number =	 11,
  pages =	 {e27533},
  issn =	 {1932-6203},
  doi =		 {10.1371/journal.pone.0027533},
}

@article{Norton2014,
  author =	 {Norton, Brianna L. and Person, Anna K. and Castillo,
                  Catherine and Pastrana, Christopher and Subramanian,
                  Melanie and Stout, Jason E.},
  title =	 {Barriers to {Using} {Text} {Message} {Appointment}
                  {Reminders} in an {HIV} {Clinic}},
  journaltitle = {Telemedicine and e-Health},
  date =	 2014,
  volume =	 20,
  number =	 1,
  pages =	 {86--89},
  issn =	 {1530-5627},
  doi =		 {10.1089/tmj.2012.0275},
}

@article{Smillie2014,
  author =	 {Smillie, Kirsten and Van Borek, Natasha and Abaki,
                  Joshua and Pick, Neora and Maan, Evelyn J. and
                  Friesen, Karen and Graham, Rebecca and Levine, Sarah
                  and van der Kop, Mia L. and Lester, Richard T. and
                  Murray, Melanie},
  title =	 {A {Qualitative} {Study} {Investigating} the {Use} of
                  a {Mobile} {Phone} {Short} {Message} {Service}
                  {Designed} to {Improve} {HIV} {Adherence} and
                  {Retention} in {Care} in {Canada} ({WelTel} {BC1})},
  journaltitle = {Journal of the Association of Nurses in AIDS Care},
  date =	 2014,
  volume =	 25,
  number =	 6,
  pages =	 614,
  issn =	 {1055-3290},
  doi =		 {10.1016/j.jana.2014.01.008},
}

@article{Costa2012,
  author =	 {{da Costa}, Thiago Martini and Barbosa, Bárbara
                  Jaqueline Peres and e Costa, Durval Alex Gomes and
                  Sigulem, Daniel and de Fátima Marin, Heimar and
                  Filho, Adauto Castelo and Pisa, Ivan Torres},
  title =	 {Results of a randomized controlled trial to assess
                  the effects of a mobile SMS-based intervention on
                  treatment adherence in HIV/AIDS-infected Brazilian
                  women and impressions and satisfaction with respect
                  to incoming messages},
  journaltitle = {International Journal of Medical Informatics},
  date =	 2012,
  volume =	 81,
  number =	 4,
  pages =	 {257--269},
  issn =	 {1386-5056},
  doi =		 {10.1016/j.ijmedinf.2011.10.002},
}

@article{Mayer2017,
  title =	 {Meta-analysis on the effect of text message
                  reminders for {HIV}-related compliance},
  volume =	 29,
  issn =	 {0954-0121, 1360-0451},
  doi =		 {10.1080/09540121.2016.1214674},
  number =	 4,
  journaltitle = {AIDS Care},
  author =	 {Mayer, Jonathan E. and Fontelo, Paul},
  date =	 2017,
  pages =	 {409--417},
}

@article{Sabin2015,
  author =	 {Sabin, Lora L. and Bachman DeSilva, Mary and Gill,
                  Christopher J. and Zhong, Li and Vian, Taryn and
                  Xie, Wubin and Cheng, Feng and Xu, Keyi and Lan,
                  Guanghua and Haberer, Jessica E. and Bangsberg,
                  David R. and Li, Yongzhen and Lu, Hongyan and
                  Gifford, Allen L.},
  title =	 {Improving Adherence to Antiretroviral Therapy With
                  Triggered Real-time Text Message Reminders: The
                  China Adherence Through Technology Study},
  journaltitle = {JAIDS Journal of Acquired Immune Deficiency
                  Syndromes},
  date =	 2015,
  volume =	 69,
  number =	 5,
  pages =	 551,
  issn =	 {1525-4135},
  doi =		 {10.1097/QAI.0000000000000651},
}

@article{Moore2015,
  author =	 {Moore, David J. and Poquette, Amelia and Casaletto,
                  Kaitlin B. and Gouaux, Ben and Montoya, Jessica
                  L. and Posada, Carolina and Rooney, Alexandra S. and
                  Badiee, Jayraan and Deutsch, Reena and Letendre,
                  Scott L. and Depp, Colin A. and Grant, Igor and
                  Atkinson, J. Hampton and {The HIV Neurobehavioral
                  Research Program (HNRP) Group}},
  title =	 {Individualized Texting for Adherence Building
                  (iTAB): Improving Antiretroviral Dose Timing Among
                  HIV-Infected Persons with Co-occurring Bipolar
                  Disorder},
  journaltitle = {AIDS and Behavior},
  date =	 2015,
  volume =	 19,
  number =	 3,
  pages =	 {459--471},
  issn =	 {1573-3254},
  doi =		 {10.1007/s10461-014-0971-0},
}

@article{Muessig2015,
  author =	 {Muessig, Kathryn E. and Nekkanti, Manali and
                  Bauermeister, Jose and Bull, Sheana and
                  Hightow-Weidman, Lisa B.},
  title =	 {A Systematic Review of Recent Smartphone, Internet
                  and Web 2.0 Interventions to Address the HIV
                  Continuum of Care},
  journaltitle = {Current HIV/AIDS Reports},
  date =	 2015,
  volume =	 12,
  number =	 1,
  pages =	 {173--190},
  issn =	 {1548-3576},
  doi =		 {10.1007/s11904-014-0239-3},
}

@article{Nhavoto2017,
  author =	 {Nhavoto, José António and Grönlund, Åke and Klein,
                  Gunnar O.},
  title =	 {Mobile health treatment support intervention for HIV
                  and tuberculosis in Mozambique: Perspectives of
                  patients and healthcare workers},
  journaltitle = {PLOS ONE},
  date =	 2017,
  volume =	 12,
  number =	 4,
  pages =	 {e0176051},
  issn =	 {1932-6203},
  doi =		 {10.1371/journal.pone.0176051},
}

@article{Shah2019,
  author =	 {Shah, Reshma and Watson, Julie and Free, Caroline},
  title =	 {A systematic review and meta-analysis in the
                  effectiveness of mobile phone interventions used to
                  improve adherence to antiretroviral therapy in HIV
                  infection},
  journaltitle = {BMC Public Health},
  date =	 2019,
  volume =	 19,
  number =	 1,
  pages =	 915,
  issn =	 {1471-2458},
  doi =		 {10.1186/s12889-019-6899-6},
}

@article{Schneider2004,
  author =	 {Schneider, John and Kaplan, Sherrie H. and
                  Greenfield, Sheldon and Li, Wenjun and Wilson, Ira
                  B.},
  title =	 {Better physician-patient relationships are
                  associated with higher reported adherence to
                  antiretroviral therapy in patients with HIV
                  infection},
  journaltitle = {Journal of General Internal Medicine},
  date =	 2004,
  volume =	 19,
  number =	 11,
  pages =	 {1096--1103},
  issn =	 {0884-8734, 1525-1497},
  doi =		 {10.1111/j.1525-1497.2004.30418.x},
}

@article{Martin2005,
  author =	 {Martin, Leslie R. and Williams, Summer L. and
                  Haskard, Kelly B. and DiMatteo, M Robin },
  title =	 {The challenge of patient adherence},
  journaltitle = {Therapeutics and Clinical Risk Management},
  date =	 2005,
  volume =	 1,
  number =	 3,
  pages =	 {189--199},
  doi =		 {10.2147/tcrm.s12160382},
}

@article{Dowshen2012,
  author =	 {Dowshen, Nadia and Kuhns, Lisa M. and Johnson, Amy
                  and Holoyda, Brian James and Garofalo, Robert},
  title =	 {Improving Adherence to Antiretroviral Therapy for
                  Youth Living with HIV/AIDS: A Pilot Study Using
                  Personalized, Interactive, Daily Text Message
                  Reminders},
  journaltitle = {Journal of Medical Internet Research},
  date =	 2012,
  volume =	 14,
  number =	 2,
  pages =	 {e2015},
  doi =		 {10.2196/jmir.2015},
}

@article{De2003,
  author =	 {De Geest, Sabina and Sabaté, Eduardo},
  title =	 {Adherence to Long-Term Therapies: Evidence for
                  Action},
  journaltitle = {European Journal of Cardiovascular Nursing},
  date =	 2003,
  volume =	 2,
  number =	 4,
  pages =	 {323--323},
  issn =	 {1474-5151, 1873-1953},
  doi =		 {10.1016/s1474-5151(03)00091-4},
}

@article{Osterberg2005,
  author =	 {Osterberg, Lars and Blaschke, Terrence},
  title =	 {Adherence to Medication},
  journaltitle = {New England Journal of Medicine},
  date =	 2005,
  volume =	 353,
  number =	 5,
  pages =	 {487--497},
  issn =	 {0028-4793, 1533-4406},
  doi =		 {10.1056/NEJMra050100},
}

@article{Armitage2020,
  author =	 {Armitage, Laura Catherine and Kassavou, Aikaterini
                  and Sutton, Stephen},
  title =	 {Do mobile device apps designed to support medication
                  adherence demonstrate efficacy? A systematic review
                  of randomised controlled trials, with meta-analysis},
  journaltitle = {BMJ Open},
  date =	 2020,
  volume =	 10,
  number =	 1,
  pages =	 {e032045},
  issn =	 {2044-6055, 2044-6055},
  doi =		 {10.1136/bmjopen-2019-032045},
}

@article{Siedner2012,
  author =	 {Siedner, Mark J. and Haberer, Jessica E. and Bwana,
                  Mwebesa Bosco and Ware, Norma C. and Bangsberg,
                  David R.},
  title =	 {High acceptability for cell phone text messages to
                  improve communication of laboratory results with
                  HIV-infected patients in rural Uganda: a
                  cross-sectional survey study},
  journaltitle = {BMC Medical Informatics and Decision Making},
  date =	 2012,
  volume =	 12,
  number =	 1,
  pages =	 56,
  issn =	 {1472-6947},
  doi =		 {10.1186/1472-6947-12-56},
}

@article{Orr2015,
  author =	 {Orr, Jayne A. and King, Robert J.},
  title =	 {Mobile phone SMS messages can enhance healthy
                  behaviour: a meta-analysis of randomised controlled
                  trials},
  journaltitle = {Health Psychology Review},
  date =	 2015,
  volume =	 9,
  number =	 4,
  pages =	 {397--416},
  issn =	 {1743-7199, 1743-7202},
  doi =		 {10.1080/17437199.2015.1022847},
}

@article{Hirsch-Moverman2017,
  author =	 {Hirsch-Moverman, Yael and Daftary, Amrita and
                  Yuengling, Katharine A. and Saito, Suzue and Ntoane,
                  Moeketsi and Frederix, Koen and Maama, Llang B. and
                  Howard, Andrea A.},
  title =	 {Using mHealth for HIV/TB Treatment Support in
                  Lesotho: Enhancing Patient–Provider Communication in
                  the START Study},
  journaltitle = {JAIDS Journal of Acquired Immune Deficiency
                  Syndromes},
  date =	 2017,
  volume =	 74,
  pages =	 {S37},
  issn =	 {1525-4135},
  doi =		 {10.1097/QAI.0000000000001202},
}

@article{Chib2012,
  author =	 {Chib, Arul and Wilkin, Holley and Ling, Leow Xue and
                  Hoefman, Bas and Van Biejma, Hajo},
  title =	 {You Have an Important Message! Evaluating the
                  Effectiveness of a Text Message HIV/AIDS Campaign in
                  Northwest Uganda},
  journaltitle = {Journal of Health Communication},
  date =	 2012,
  volume =	 17,
  number =	 {sup1},
  pages =	 {146--157},
  issn =	 {1081-0730, 1087-0415},
  doi =		 {10.1080/10810730.2011.649104},
  urldate =	 {2024-02-25},
}

@article{Whiteley2021,
  author =	 {Whiteley, Laura and Olsen, Elizabeth M. and
                  Haubrick, Kayla K. and Odoom, Enyonam and Tarantino,
                  Nicholas and Brown, Larry K.},
  title =	 {A Review of Interventions to Enhance HIV Medication
                  Adherence},
  journaltitle = {Current HIV/AIDS Reports},
  date =	 2021,
  volume =	 18,
  number =	 5,
  pages =	 {443--457},
  issn =	 {1548-3576},
  doi =		 {10.1007/s11904-021-00568-9},
}

@article{Spaan2020,
  author =	 {Spaan, Pascalle and van Luenen, Sanne and Garnefski,
                  Nadia and Kraaij, Vivian},
  title =	 {Psychosocial interventions enhance HIV medication
                  adherence: A systematic review and meta-analysis},
  journaltitle = {Journal of Health Psychology},
  date =	 2020,
  volume =	 25,
  number =	 {10-11},
  pages =	 {1326--1340},
  issn =	 {1359-1053, 1461-7277},
  doi =		 {10.1177/1359105318755545},
}

@article{Kalichman2017,
  author =	 {Kalichman, Seth C. and Kalichman, Moira O. and
                  Cherry, Chauncey},
  title =	 {Forget about forgetting: structural barriers and
                  severe non-adherence to antiretroviral therapy},
  journaltitle = {AIDS Care},
  date =	 2017,
  volume =	 29,
  number =	 4,
  pages =	 {418--422},
  issn =	 {0954-0121, 1360-0451},
  doi =		 {10.1080/09540121.2016.1220478},
}

@article{Ondenge2017,
  author =	 {Ondenge, Ken and Renju, Jenny and Bonnington, Oliver
                  and Moshabela, Mosa and Wamoyi, Joyce and Nyamukapa,
                  Constance and Seeley, Janet and Wringe, Alison and
                  Skovdal, Morten},
  title =	 {‘I am treated well if I adhere to my HIV
                  medication’: putting patient–provider interactions
                  in context through insights from qualitative
                  research in five sub-Saharan African countries},
  journaltitle = {Sexually Transmitted Infections},
  date =	 2017,
  volume =	 93,
  number =	 {Suppl 3},
  issn =	 {1368-4973, 1472-3263},
  doi =		 {10.1136/sextrans-2016-052973},
  pmid =	 28736392,
}

@article{Haskard2009,
  author =	 {Haskard Zolnierek, Kelly B. and DiMatteo, M. Robin},
  title =	 {Physician Communication and Patient Adherence to
                  Treatment: A Meta-Analysis},
  journaltitle = {Medical Care},
  date =	 2009,
  volume =	 47,
  number =	 8,
  pages =	 826,
  issn =	 {0025-7079},
  doi =		 {10.1097/MLR.0b013e31819a5acc},
}

@Report{UNADIS2022,
  author =	 {{UNAIDS}},
  title =	 {In Danger: UNAIDS Global AIDS Update 2022.},
  date =	 2022,
  pagetotal =	 376,
  url =
                  {https://www.unaids.org/sites/default/files/media_asset/2022-global-aids-update_en.pdf},
}

\section*{Agradecimientos}
\label{sec:org794192b}
Nos gustaría agradecer a varias personas el apoyo recibido durante este estudio. Estamos profundamente agradecidos a los entrevistados (partidarios del tratamiento del VIH, personal de las clínicas de TAR, clientes expertos, educadores sanitarios, funcionarios de OpenMRS, consejeros de VIH, equipos de salud de las aldeas y trabajadores sanitarios) que ofrecieron libremente su valioso tiempo para ser entrevistados --- muchas gracias por proporcionarnos toda la valiosa información necesaria para esta investigación.

\section*{Biografías}
\label{sec:orgfe41562}
Livingstone Njuba es becario de posgrado de Equity \& Merit Scholarship en la Universidad de Manchester (Reino Unido). Realizó un máster en Gestión y Sistemas de Información: Cambio y Desarrollo en la Universidad de Manchester, una licenciatura en Informática en la Universidad Makerere de Kampala (Uganda), y un diploma de posgrado en Gestión Empresarial en la Universidad Amity de la India. Actualmente es Director de TIC en Kalangala Infrastructure Services Limited, una empresa privada de servicios múltiples que proporciona transporte por transbordador, electricidad y agua limpia en la isla de Bugala, distrito de Kalangala (Uganda). Sus intereses de investigación actuales se centran en las TIC para el desarrollo, la sanidad móvil, los SIG para servicios públicos y las tecnologías de redes inteligentes.

El Dr. Juan E. Gómez-Morantes es profesor asistente en la Facultad de Ingeniería de la Pontificia Universidad Javeriana de Bogotá (Colombia). Realizó un máster en Ingeniería en la Universidad de los Andes de Bogotá (Colombia) y un doctorado en Política y Gestión del Desarrollo en la Universidad de Manchester, con una tesis sobre los procesos de innovación digital en microempresas vulnerables. Además de este tema, sus intereses de investigación incluyen las TIC para el desarrollo, los aspectos sociales de la tecnología, las políticas públicas de TIC, y la intersección entre las tecnologías digitales y el calentamiento global.

La Dra. Andrea Herrera es Profesora Adjunta del Departamento de Ingeniería de Sistemas y Computación de la Facultad de Ingeniería de la Universidad de los Andes en Bogotá, Colombia. Su investigación se centra en cómo las tecnologías de la información y la comunicación (TIC) pueden hacer que las personas, las organizaciones, y las cadenas de suministro sean más resilientes. La Dra. Herrera ha actuado como revisora en conferencias sobre sistemas de información (por ejemplo, AMCIS y PACIS) y es una entusiasta seguidora del uso de las TIC, principalmente las redes sociales, para la gestión de emergencias, y el análisis de big data.

La Dra. Sonia Camacho es Profesora Asociada de la Facultad de Administración de la Universidad de los Andes en Bogotá, Colombia. También es la Directora del Comité de Investigación de la Facultad desde 2020. Su área de investigación se enmarca en el ámbito de la interacción persona-computador. Sus intereses de investigación incluyen el lado oscuro de la tecnología de la información, el comercio electrónico, y los impactos positivos del uso de la tecnología de la información. Tiene un doctorado en Administración de Empresas de la Universidad McMaster (Hamilton, Ontario), una maestría en Administración (Universidad de los Andes), y una licenciatura en Ingeniería de Sistemas (Universidad Nacional de Colombia).
\end{document}